\documentclass[aps,prx,reprint,preprintnumbers,superscriptaddress,nofootinbib,longbibliography,floatfix]{revtex4-1}

\pdfoutput=1

\usepackage{graphicx, subcaption}
\usepackage{dcolumn}
\usepackage{bm}
\usepackage{xcolor}
\usepackage{amsmath}
\usepackage{multirow}
\usepackage{caption}
\usepackage{float}
\usepackage{tabularx}

\usepackage{hyperref}
\hypersetup{
  colorlinks=true,
  citecolor=blue,
  linkcolor=blue,
  urlcolor=blue
}

\DeclareRobustCommand{\Sec}[1]{Sec.~\ref{#1}}

\DeclareRobustCommand{\App}[1]{App.~\ref{#1}}
\DeclareRobustCommand{\Tab}[1]{Table~\ref{#1}}

\DeclareRobustCommand{\Fig}[1]{Fig.~\ref{#1}}
\DeclareRobustCommand{\Figs}[2]{Figs.~\ref{#1} and \ref{#2}}

\begin{document}


\title{FETA: Flow-Enhanced Transportation for Anomaly Detection}

\author{Tobias Golling}
\email{tobias.golling@unige.ch}
\affiliation{Département de Physique Nucléaire et Corpusculaire, Université de Genève, Genève; Switzerland}

\author{Samuel Klein}
\email{samuel.klein@unige.ch}
\affiliation{Département de Physique Nucléaire et Corpusculaire, Université de Genève, Genève; Switzerland}

\author{Radha Mastandrea}
\email{rmastand@berkeley.edu}
\affiliation{Department of Physics, University of California, Berkeley, CA 94720, USA}
\affiliation{Physics Division, Lawrence Berkeley National Laboratory, Berkeley, CA 94720, USA}

\author{Benjamin Nachman}
\email{bpnachman@lbl.gov}
\affiliation{Physics Division, Lawrence Berkeley National Laboratory, Berkeley, CA 94720, USA}
\affiliation{Berkeley Institute for Data Science, University of California, Berkeley, CA 94720, USA}

\date{\today}

\begin{abstract}

Resonant anomaly detection is a promising framework for model-independent searches for new particles. Weakly supervised resonant anomaly detection methods compare data with a potential signal against a template of the Standard Model (SM) background inferred from sideband regions. We propose a means to generate this background template that uses a flow-based model to create a mapping between high-fidelity SM simulations and the data. The flow is trained in sideband regions with the signal region blinded, and the flow is conditioned on the resonant feature (mass) such that it can be interpolated into the signal region.  To illustrate this approach, we use simulated collisions from the Large Hadron Collider (LHC) Olympics Dataset.  We find that our flow-constructed background method has competitive sensitivity with other recent proposals and can therefore provide complementary information to improve future searches.

\end{abstract}

\maketitle


\section{Introduction}
\label{sec:level1}

While the Large Hadron Collider (LHC) has been operational for over a decade, no targeted search for physics beyond the Standard Model has found significant evidence for new particles or forces of nature. As a result, there has recently been a growing interest in \textit{model-agnostic} searches: in such studies, the goal is to find anomalies while making few assumptions about the underlying physics model that could have produced them. This mode of analysis is highly general, and it has gained momentum from recent advances for modeling, classifying, and finding anomalies in data. For a broad review of the role of modern machine learning (ML) methods in searches for new physics, see \cite{Karagiorgi:2022qnh}; for performance summaries of ML-inspired methods on LHC-like data, see the LHC Olympics \cite{Kasieczka:2021xcg} and Dark Machines Anomaly Score Challenge \cite{Aarrestad:2021oeb} reports. 

One class of anomaly detection searches is for \textit{resonant} anomalies. In this case, we assume that the beyond-the-Standard Model (BSM) physics signature would be resonant in at least one feature, typically some sort of mass. Such a signal might correspond to the production of a new on-shell particle. These ``bump hunts" search for an excess in the mass feature above the known Standard Model (SM) background in a well-defined signal region. However, such searches require the existence of accurate SM background templates within the signal region. While particle generators for SM physics exist, the resulting simulated data (even with detector effects added) is not accurate enough to be used as a background for bump hunts in hadronic final states: hard process generators carry out perturbative calculations only to NLO, showering simulators make use of heuristic nonperturbative models, and detector simulators make necessary simplifying assumptions about particle-detector interactions in order to speed up runtime.

In response to this problem of subaccurate particle generators, a number of studies have focused on constructing SM background templates for LHC-like detected data within a signal region. Broadly, methods can be categorized on two qualities:

\begin{enumerate}
    \item \textit{Simulation-assisted} vs. \textit{data-driven}. For simulation assisted methods, the background template construction is informed by a set of simulated LHC-like collider data representing SM processes; for data-driven methods, data from sidebands mass regions is used, where the sidebands are far enough from the resonant process such that the data can be treated as a proxy for SM background.
    \item \textit{Likelihood learning} vs.  \textit{feature morphing}. For the former, methods learn the likelihood of a SM-only dataset (such as simulation, or detected data in sidebands) where there are no signal events. This learned likelihood is then interpolated into the signal region to act as a SM background template above which signal events might be detected. Alternatively, methods can physically morph features from the SM-only dataset to the detected, signal-containing dataset.
\end{enumerate}

A number of previous methods for SM background construction are classified visually in \Tab{tab:methods}. This two-axis scheme is not meant to be all-encompassing, as there exist many methods for anomaly detection that cannot be so neatly classified (see \cite{Benkendorfer:2020gek, Amram_2021, Choi:2020bnf, CMS:2022uga} for examples of such methods in practice).

\begin{table}[]
\footnotesize
    \centering
    \begin{tabular}{|c|c|c|}
    \hline
      & Simulation-assisted  & Data-driven  \\
      \hline
      \parbox[][30pt][c]{1.7cm}{Likelihood learning} & \textsc{Salad} \cite{Andreassen:2020nkr}  &  \parbox{3.5cm}{Overdensity searches \cite{Stein:2020rou},\\\textsc{Anode} \cite{Nachman:2020lpy}, \\\textsc{(La)Cathode}  \cite{Hallin:2021wme, Hallin:2022eoq}}  \\
      \hline
      \parbox[][20pt][c]{1.7cm}{Feature morphing}    & \textsc{Feta} (this work) & \textsc{Curtains} \cite{Raine:2022hht} \\
      \hline
    \end{tabular}
    \caption{Broadly speaking, many methods for constructing SM background templates for resonant anomaly detection can be classified on two axes. In this study, we introduce the \textsc{Feta} method.}
    \label{tab:methods}
\end{table}

In this paper, we propose the ``Flow-Enhanced Transportation for Anomaly Detection" (\textsc{Feta}) method to model SM-like background at the LHC. \textsc{Feta} is simulation-assisted and relies on feature morphing, therefore filling the previously empty section of \Tab{tab:methods}.

In \textsc{Feta}, we train a normalizing flow\footnote{A comprehensive review of flow-based models is given in \cite{Kobyzev_2021}.} \cite{tabak_flows} to learn a mapping between simulation and data in sidebands regions. We then apply the learned mapping to signal region simulation to create a simulation-informed template for signal region SM background. \textsc{Feta} benefits from being simulation-assisted since it can use simulated SM data as a physically-informed prior for the background template; the method further benefits from using feature morphing since it is robust to mapping between feature spaces with non-overlapping support.

The structure of this paper is as follows. In \Sec{sec:method}, we provide a concise background of normalizing flows and outline how they will be applied to physics-specific datasets. In \Sec{sec:toy_model}, we illustrate the effectiveness of flow-based models for creating context-dependent mappings with a toy example of triangular datasets. In \Sec{sec:lhc_data}, we exchange the toy models for LHC-like data and use \textsc{Feta} to create a model for LHC-like detected SM data. In \Sec{sec:anomaly_detection}, we test the performance of \textsc{Feta} in a series of realistic anomaly detection tasks. In \Sec{sec:conclusions}, we conclude and suggest avenues for further study.

\section{Methodology}
\label{sec:method}

\subsection{Normalizing flows as morphing functions}

Normalizing flows are constructed from invertible neural networks between sets of variables sampled from different probability densities. Given a random variable $X$ sampled from a \textit{reference distribution} $p_R$, one can define a transformation $T$ that produces another random variable $Z$, i.e. $Z = T(X)$. The density of $Z$ is then given by $p_Z(Z) =  p_R(X)|\textrm{det}\frac{\partial T}{\partial X}|^{-1}$. By chaining together a number of transformations $T_i$, one can produce an arbitrarily complex mapping between the initial reference density $p_R$ and a \textit{target distribution} $p_T$. Typically, the target density is taken to be a standard normal distribution. (Hence the name ``normalizing" flow.)

In this work, we use the normalizing flow both for its density estimation power and its ability to construct morphing functions between nontrivial reference $X_R \sim p_R$ and target $X_T \sim p_T$ distributions. Our reference density is derived from Standard Model simulation ($X_\mathrm{SIM} \sim p_\mathrm{SIM}$), and our target dataset is derived from detected data ($X_\mathrm{DAT} \sim p_\mathrm{DAT}$).  

\begin{figure}
    \centering
    \includegraphics[width=\linewidth]{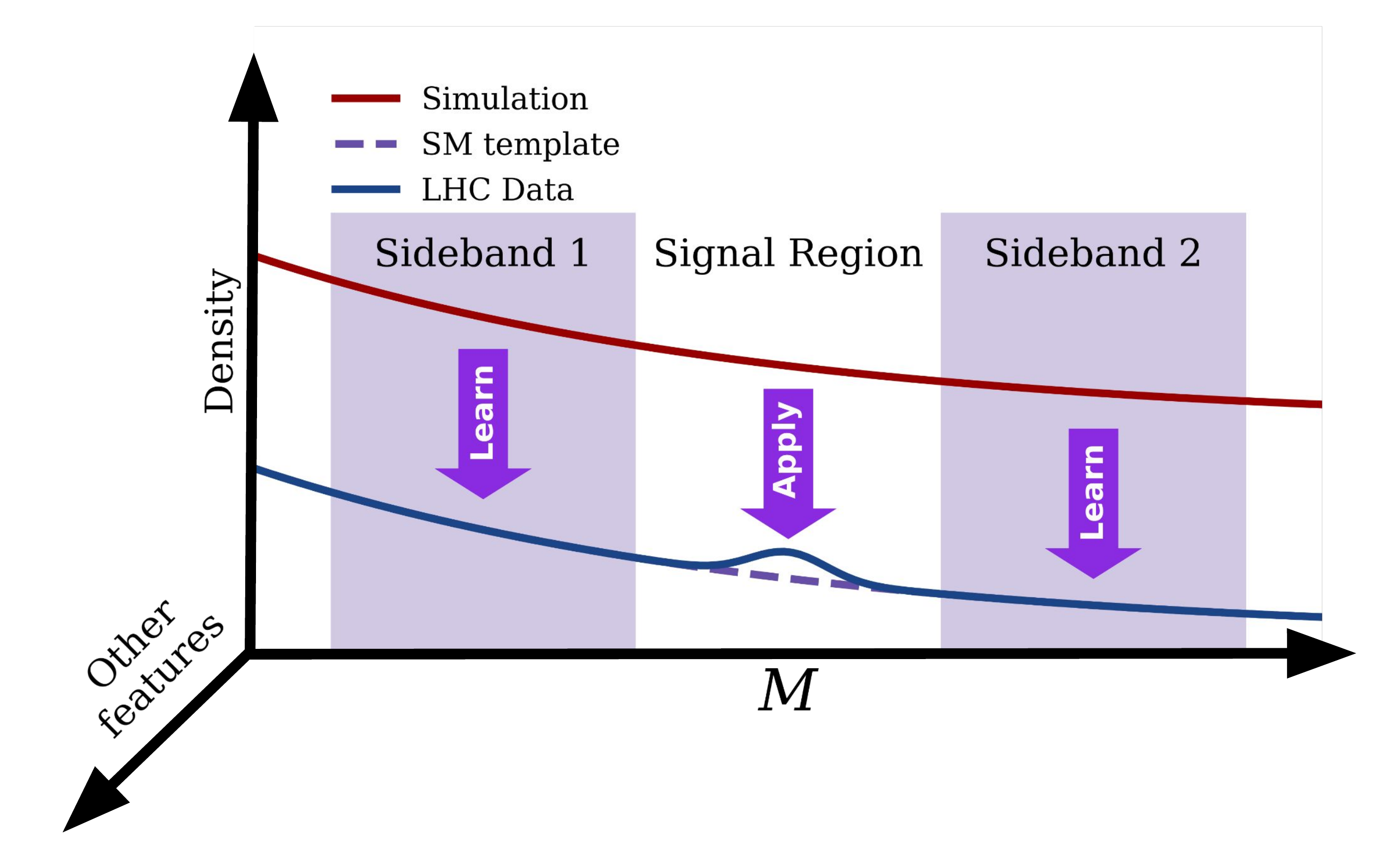}
    \caption{A schematic of the \textsc{Feta} method. We train a flow to learn the mapping between simulation and data in sidebands regions, which are expected to be background-only. We then apply the learned flow to simulation in a signal region to produce an approximation for background in that region.}
    \label{fig:schematic}
\end{figure}

More explicitly, we define a set of $N$ event observables such that events in the reference and target are $N$-dimensional vectors $X_\mathrm{SIM}$ and $X_\mathrm{DAT}$, which respectively are sampled from the $N$-dimensional feature densities $p_\mathrm{SIM}$ and $p_\mathrm{DAT}$. We then train a flow to learn the mapping $T$ between $p_\mathrm{SIM}$ and $p_\mathrm{DAT}$. In training the flow, we must ensure that the learned mapping is between simulated Standard-Model background and LHC-detected \textit{background}. For resonant anomalies, we assume that the signal will be localized in one feature $m_{res}$. This allows us to define sidebands (SB) and a signal region (SR) in $m_{res}$, where data from the former regions is assumed to follow the SM distribution. We then train the flow only on data from SB, using the resonant feature to \textit{condition} the mapping $T(\cdot|m_{res})$. The learned flow is then applied to simulation in the SR to produce an approximation of $X_\textrm{DAT}$: $X^*_\textrm{SIM} = T(X_\textrm{SIM}|m_{res})$ for LHC-detected background in the SR. A schematic of this method is shown in \Fig{fig:schematic}.

There are several advantages to using a flow-based architecture over other architectures such as GANs or VAEs. Normalizing flows are known to be more stable and achieve convergence during training faster, especially in higher dimensions. This property allows for the freedom to choose a larger feature space $X$, which may be desirable for a model-agnostic study. For GANs specifically, attempting to learn conditional mappings between datasets is not an easily done task. In addition, the density estimating power of the normalizing flow allows us to \textit{oversample} from the reference distribution and reduce statistical uncertainties (explored in more detail in \App{sec:app_oversampling}). This is not possible with VAEs, which require the definition of explicit pairs to train the encoder-decoder architecture.

\subsection{\label{sec:data}Flow construction}
All flows were constructed using the \texttt{nflows} package \cite{nflows} and were trained using \textsc{PyTorch} \cite{NEURIPS2019_9015}.

As an important procedural note: to learn a flow that maps between two nontrivial densities, we use a two-step procedure. We first train a \textit{Base Density} normalizing flow to learn the mapping between a normal distribution and the reference density $p_R$ across all mass bands (i.e. using reference data from both the SB and SR). We then train a \textit{Transport} flow to learn the mapping between the  \textit{Base Density} distribution and the target density $p_T$, this time only using data from the SB. This specific method, which allows for the use of flows to map between nontrivial distributions, was proposed and implemented in the \textsc{Curtains} background construction method (and is further explored in \cite{https://doi.org/10.48550/arxiv.2211.02487}). It is thus more accurate to simply call \textsc{Feta} a flow-based method, rather than a normalizing flow-based one.

\section{Testing the \textsc{Feta} method with a toy model}
\label{sec:toy_model}

To concretely illustrate the \textsc{Feta} method, we train a flow to map between two toy triangular datasets. Both datasets contain 100,000 samples in a two-dimensional feature space, with a feature $X$ that is conditionally dependent on a feature $M$. This conditioning feature can be interpreted as a mass-like feature, or one in which we expect an anomaly to be resonant.

\subsection{Toy model datasets}

To construct each dataset, we first generate samples of the conditioning (resonant) feature $M$, which we take to be uniformly distributed between (0, 1). On this feature, we define the signal region SR $\sim$ (0.34, 0.66). We also define two sidebands, SB1 $\sim$ (0.0, 0.34) on the low mass side of the SR, and SB2 $\sim$ (0.66, 1.0) on the high mass side.

\begin{figure}
    \centering
    \includegraphics[width = \linewidth]{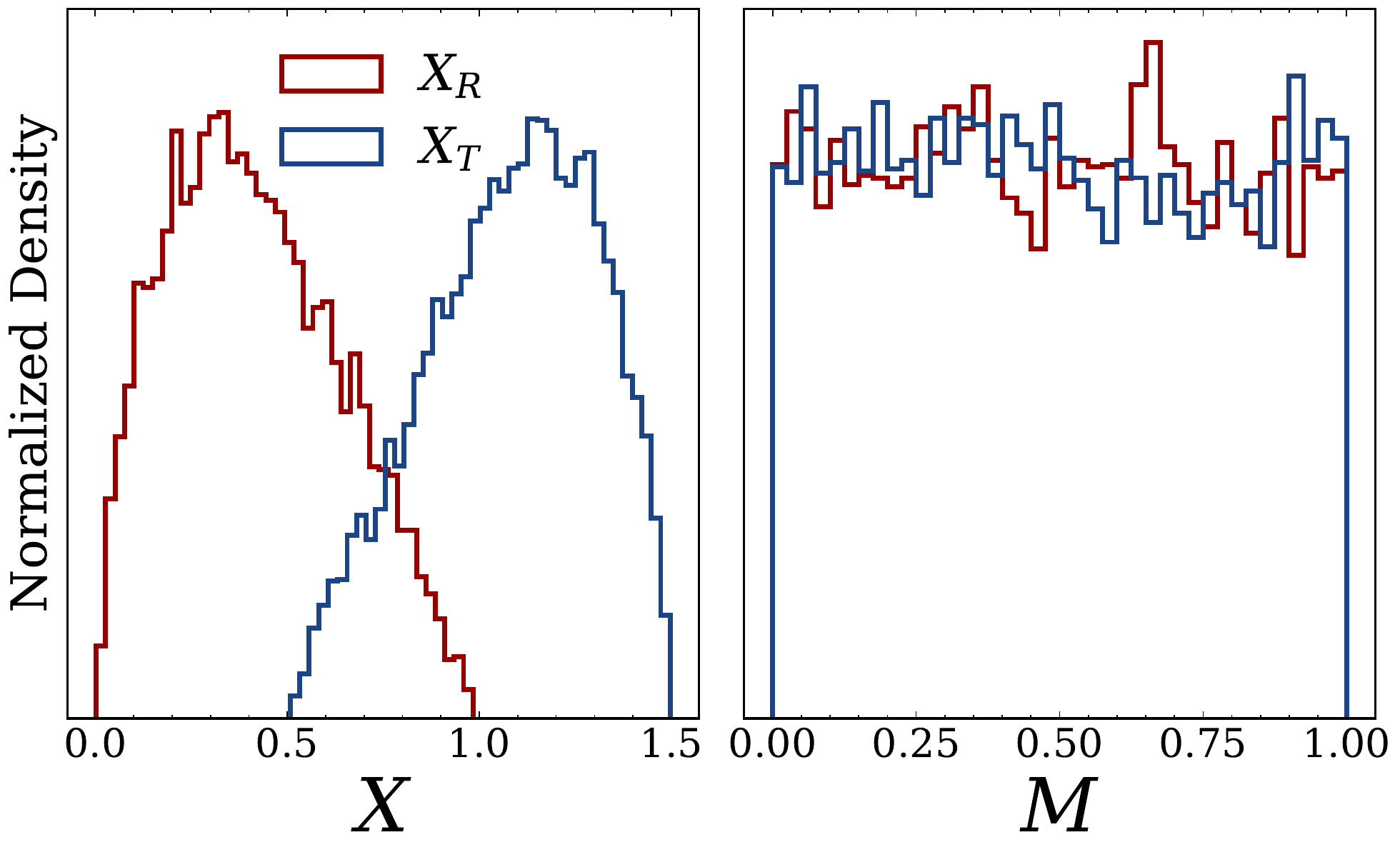}
    \caption{Toy datasets $X_\textrm{R}$ and $X_\textrm{T}$. The feature $X$ is a function of the conditioning feature $M$.}
    \label{fig:toy_model_datasets}
\end{figure}

For the corresponding nonresonant feature, we draw samples from a triangular dataset with endpoints at (0, 1). In order to condition this feature $X$ on the mass, for each sample we define the midpoints $m_R$, $m_T$ of the triangular distributions for reference and target to be linear functions of $M$, 
$m_\textrm{R} = 0.4995M$ and $m_\textrm{T} = 0.4995M + 0.5005$. Then the toy reference dataset $X_R$ consists of left-triangular dataset samples in the range (0, 1), and the toy target dataset $X_T$ consists of right-triangular samples. We further shift all the samples $X_T$ of the target dataset by 0.5 such that the reference and the target sets have nonidentical support. Samples from the two datasets are shown in \Fig{fig:toy_model_datasets}.

The flow architecture and hyperparameters are given in \Tab{tab:toy_model_flow}. All settings were optimized via manual tuning and were chosen to give the best possible performance on the SB regions (as quantified by the ROC AUC, which is defined in \Sec{sec:toy_model_results}). Flow training is optimized with \textsc{AdamW} \cite{adam}, and the learning rate is annealed to zero following a cosine schedule \cite{DBLP:journals/corr/LoshchilovH16a}. Before training, all features are minmaxscaled to the range (-3, 3), which was found to be optimal with respect to the flow training; further, the samples are split into training (80\%) and validation (20\%) sets. The model from the epoch with the lowest validation loss is used for evaluation.

\begin{table}[]
    \centering
    \begin{tabular}{|c|c|c|}
    \hline
    \hline
    Parameter & \textit{Base density} flow & \textit{Transport} flow  \\
    \hline
    \hline
     Flow type & Autoregressive \cite{https://doi.org/10.48550/arxiv.1804.00779} & Coupling \\
     Spline & Piecewise RQ & Piecewise RQ \\
      Num. MADE  blocks  & 8  & 8  \\
     Num. layers & 2 & 2 \\
     Num. hidden features & 16 & 16 \\
      \hline
        Epochs & 20 & 20  \\
       Batch size & 128 & 256 \\
       Learning rate & 3$\times 10^{-4}$ & 3$\times 10^{-4}$ \\
       Weight Decay & 1$\times 10^{-4}$ & 1$\times 10^{-5}$ \\
       \hline
       \hline
    \end{tabular}
    \caption{(RQ = rational quadratic; MADE = Masked Autoencoder for Distribution Estimation \cite{https://doi.org/10.48550/arxiv.1502.03509}.) Flow architecture and training hyperparameters used for the toy (triangular) dataset. Both the \textit{Base Density} and \textit{Transport} flow parameters were optimized through manual tuning.}
    \label{tab:toy_model_flow}
\end{table}

\subsection{Toy model results}
\label{sec:toy_model_results}

\begin{figure*}
     \centering
     
         \includegraphics[width=.8\textwidth]{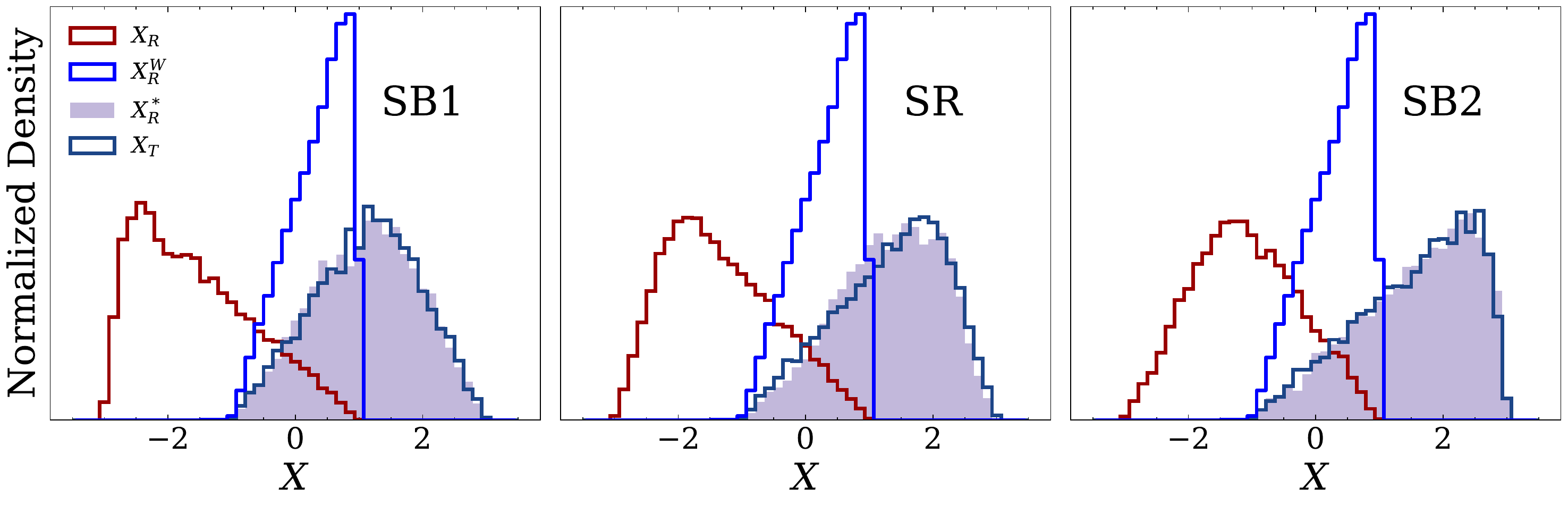}
    
        \caption{Probability distributions for $X_R$, $X^*_R$, and $X_T$ for the toy model. The $X^*_R$ samples represent the trained flow acting upon $X_R$. The good agreement between $X^*_R$ and $X_T$ indicates that the flow has successfully learned to map between the two datasets. We also provide the distribution for reweighted reference $X^W_R$ to illustrate its failure to provide an accurate model for $X_T$, as the reference and target datasets have different regions of support.}
        \label{fig:trained_toy_model}
\end{figure*}

Plots of the reference $X_R$, transformed reference $X^*_R$ as found by the \textsc{Feta} method, and target $X_T$ distributions are shown in \Fig{fig:trained_toy_model}. We compare the distributions separately in SB1, SR, and SB2 (recall that the flow is trained only on data from SB1 and SB2 and is applied to the blinded SR data). Qualitatively, there is excellent agreement between the distributions for $X^*_R$ and $X_T$ in SB1 and SB2 and good agreement in the SR. We conclude that the flow has learned to map between the reference and the target data.

In \Fig{fig:trained_toy_model}, we also plot the reweighted reference $X^W_R$ as found through the \textsc{Salad} method. For this method, we train a binary classifier to discriminate between $X_R$ and $X_T$. Such a classifier will learn the likelihood ratio between the reference and the target distributions. This likelihood ratio can then be interpolated into the SR and used to reweight $X_R$. Note that for this toy model, feature reweighting is expected to fail due to the fact that there are regions of non-overlapping support between the reference and target.

We can quantify the performance of the learned morphing function by training a 5-fold binary classifier neural network to discriminate between $X^*_R$ and $X_T$. Each classifier is a fully connected (dense) network, with linear layers of sizes (5, 64, 32, 1) and a dropout of 0.1 between each layer. Each classifier is trained for up to 100 epochs with a batch size of 128, learning rate of 10$^{-3}$, and patience of 5 epochs. We evaluate all test data on the classifier from the fold with the best (lowest) validation loss. 

As our scoring metric, we use the area under the receiver operating characteristic (ROC) curve. These AUC scores for this classifier are shown in \Tab{tab:toy_model_rocs}. An ideal morphing function would result in the transformed reference and target datasets being indistinguishable from each other, which would correspond to a AUC close to 0.50. Indeed, the AUCs hit this performance benchmark, and they are also far lower than those for a classifier trained to discriminate between untransformed $X_R$ and $X_T$.

\begin{table}[]
    \centering
    \begin{tabular}{|c|c|c|}
     \hline
     \hline
        Band & AUC for $X^*_R$ vs $X_T$ & AUC for $X_R$ vs $X_T$ \\
        \hline
        \hline
        SB1 &0.5144 $\pm$ 0.0095 & 0.9830 $\pm$ 0.0009\\
        \hline
        SR & 0.5117 $\pm$ 0.0040 & 0.9806 $\pm$ 0.0000 \\
        \hline
        SB2 & 0.5086  $\pm$ 0.0082 & 
0.9813 $\pm$ 0.0009\\
        \hline
        \hline

    \end{tabular}
    \caption{ROC AUCs for a binary classifier trained to discriminate the transformed reference $X^*_R$ from the target dataset $X_T$. For comparison, we also provide the AUCs for a binary classifier trained to discriminate the untransformed reference $X_R$ from the target. Uncertainties are the 1$\sigma$ bounds from retraining the binary classifier 20 times, each with a different random seed.}
    \label{tab:toy_model_rocs}
\end{table}

\section{Applying \textsc{Feta} to LHC-like data}
\label{sec:lhc_data}

We now move to a realistic example: training a flow to learn the morphing function between Standard Model simulation and detected data. For an ideal application of \textsc{Feta}, the reference dataset would consist of simulated Standard Model-like data ($X_R$ = $X_\mathrm{SIM}$), and the target dataset would consist of LHC-detected Standard Model data $X_T$ = $X_\mathrm{DAT}$). However, we do not have access to LHC data. Therefore in this study we use two distinct sets of simulated data for the reference and target datasets.

\subsection{LHCO datasets}

We focus on the LHC 2020 Olympics R\&D dataset \cite{LHCOlympics,Kasieczka:2021xcg}. The full dataset consists of 1,000,000 background dijet events (Standard
Model Quantum Chromodynamic (QCD) dijets) and 100,000 signal dijet events.
The signal comes from the process $Z'\rightarrow X(\rightarrow
q\overline{q})Y(\rightarrow q\overline{q})$, with three new resonances $Z'$ ($3.5$~TeV), $X$ (500~GeV),
and $Y$ (100~GeV).
Events are required to have at least one large-radius jet ($R=1$) trigger with a $p_T$ threshold of $1.2$ TeV.  The events are generated with \textsc{Herwig}++\cite{B_hr_2008}, \textsc{Pythia}~8.219~\cite{Sjostrand:2006za,Sjostrand:2014zea}, and \textsc{Delphes}~3.4.1~\cite{deFavereau:2013fsa}. Each event contains up to 700 particles with three degrees of freedom (DoF) $p_T$, $\eta$, $\phi$.

For this study, we choose a feature space of six dijet observables $m_{J_1}$, $\Delta m_{JJ}$, $\tau^{21}_{J_1}$, $\tau^{21}_{J_2}$, $\Delta R_{JJ}$, and $m_{JJ}$. Following the example of the \textsc{Curtains} analysis, we include the $\Delta R_{JJ}$ dijet observable as a feature that is highly correlated with $m_{JJ}$.

As with many anomaly detection searches, we assume that the anomaly is resonant in one feature, $m_{JJ}$. This would correspond to a new resonant particle that would produce the two jets. The assumption of resonance allows us to define a signal region (SR) in $m_{JJ}$-space, as well as two sidebands (SB1 and SB2) on the low- and high-mass ends of the SR. The region edges are defined in \Tab{tab:bands_cuts_LHC}. While we have chosen to focus on one SR / SB setup, one could in practice perform a \textit{sliding bump hunt} and scan the SR across a wide range of resonant masses.

\begin{table}[]
    \centering
  \begin{tabular}{|c|c|r|r|}
    \hline
    \hline
        Band & GeV Bounds & \textsc{Herwig++} events & \textsc{Pythia} events \\
        \hline
        \hline
        SB1 & (2900, 3300) & 210767 & 212115 \\
        SR & (3300, 3700) & 121978 & 121339\\
        SB2 & (3700, 4100) & 68609 & 66646\\
        SB1 $\cup$ SB2 & -- & 279376 & 278761\\
\hline
\hline
    \end{tabular}
    \caption{Band edge definitions in $m_{JJ}$-space for the LHC Olympics datasets and corresponding event counts.}
    \label{tab:bands_cuts_LHC}
\end{table}

\begin{figure*}
    \centering
    \includegraphics[width = \linewidth]{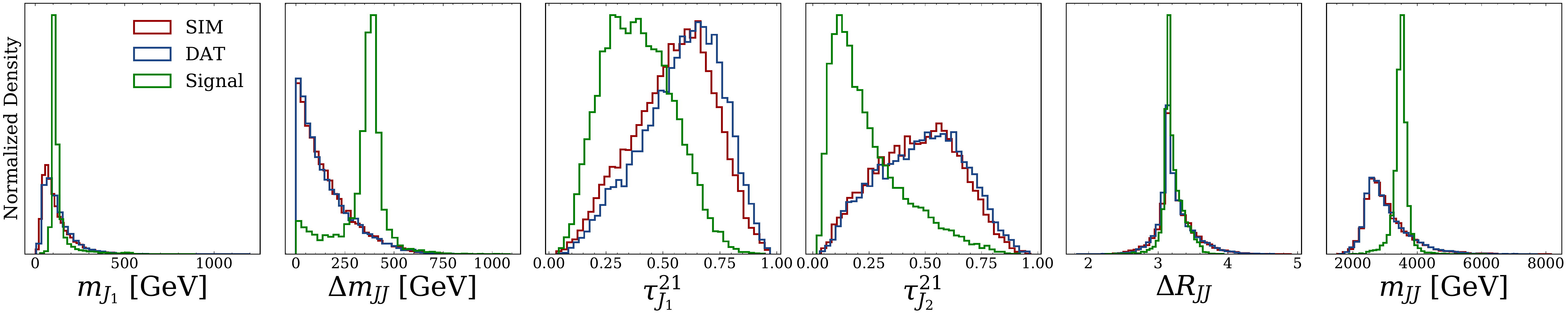}
    \caption{Feature distributions for the six dijet observables used in the LHC Olympics analysis. SIM represents \textsc{Herwig++} simulation, and DAT represents \textsc{Pythia} simulation. The last feature, $m_{JJ}$, is the feature in which we expect the anomaly to be resonant and conditions the flow mapping.}
    \label{fig:LHC_datasets}
\end{figure*}

The LHC Olympics dataset does not contain any LHC-detected data that would be the obvious choice for $X_\mathrm{SIM}$. Therefore we take  the LHC Olympics \textsc{Herwig++} data as our simulation dataset $X_\mathrm{SIM}$ and the \textsc{Pythia} data as our target dataset $X_\mathrm{DAT}$. Histograms of the six dijet observables for $X_\mathrm{SIM}$ and $X_\mathrm{DAT}$ are shown in \Fig{fig:LHC_datasets}.

\subsection{Training Method}
\label{sec:flow_training}

The method for training a flow on the LHC data is the same as for the toy dataset. However, the flow architectures used for LHC-like data are significantly more expressive. Architectures and hyperparameters are outlined in \Tab{tab:lcho_flow_params}. Notably, the \textit{Base Density} flow parameters were derived from the main architecture from \textsc{Cathode} (which relies on faithful density estimation of detected collider data in SB), and they were confirmed to give the best performance through manual tuning. Data is minmaxscaled to the range (-3, 3) before flow training, and a training-validation split of 80\%-20\% is used. All settings were manually optimized to give the best-performing flow possible, as quantified by the ROC AUCs in SB1 and SB2.

\begin{table}[]
    \centering
    \begin{tabular}{|c|c|c|}
    \hline
    \hline
    Parameter & \textit{Base Density} flow & \textit{Transport} flow  \\
    \hline
    \hline
     Flow type & Autoregressive & Coupling \\
     Spline & Piecewise RQ & Piecewise RQ \\
      Num. MADE blocks & 15  & 8  \\
     Num. layers & 1 & 2 \\
     Num. hidden features & 128 & 32 \\
      \hline
        Epochs & 100 & 50  \\
       Batch size & 128 & 256 \\
       Learning rate & 1$\times 10^{-4}$ & 5$\times 10^{-4}$ \\
       Weight Decay & 1$\times 10^{-4}$ & 1$\times 10^{-5}$ \\
       \hline
       \hline
    \end{tabular}
    \caption{Flow architecture and training hyperparameters used for SM background construction for the LHC Olympics dataset. The \textit{Base Density} flow parameters were derived from the main architecture from \textsc{Cathode}, but were confirmed to give the best performance through manual tuning. The \textit{Transport} flow parameters were optimized through manual tuning.}
    \label{tab:lcho_flow_params}
\end{table}

\subsection{LHCO results}

\begin{figure*}[hbtp]
     \centering
     \begin{subfigure}[b]{\textwidth}
         \centering
         \includegraphics[width=\textwidth]{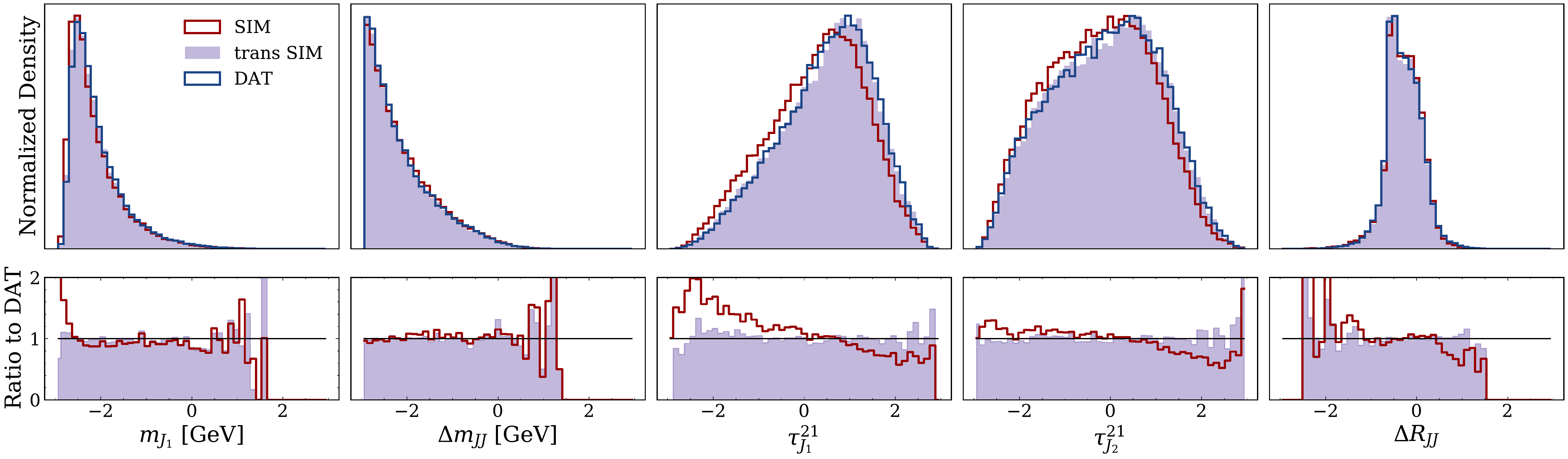}
         \caption{SB1 (low-mass sideband) distributions}
         \label{fig:LHC_results_sb1}
     \end{subfigure}
     \hfill
     \begin{subfigure}[b]{\textwidth}
         \centering
         \includegraphics[width=\textwidth]{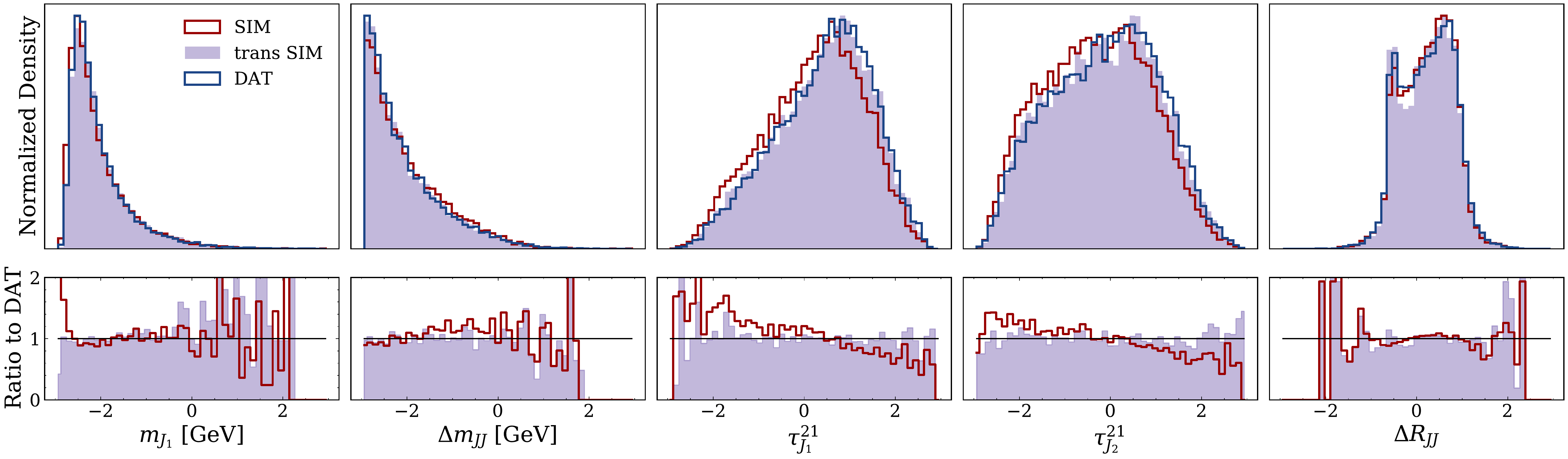}
         \caption{SB2 (high-mass sideband) distributions}
         \label{fig:LHC_results_sb2}
     \end{subfigure}
     \hfill
     \begin{subfigure}[b]{\textwidth}
         \centering
         \includegraphics[width=\textwidth]{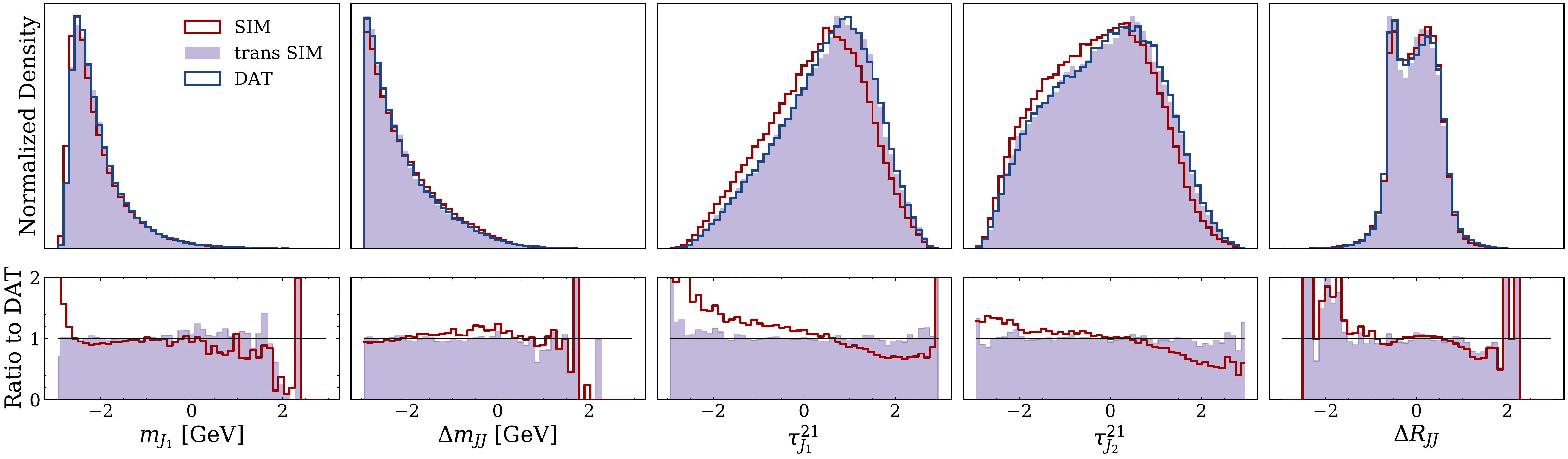}
         \caption{SR (signal region) distributions. Note that the flow was not explicitly been trained to map from $X_\mathrm{SIM}$ to $X_\mathrm{DAT}$ in this mass band, but was interpolated after training on SB1 and SB2.}
         \label{fig:LHC_results_SR}
     \end{subfigure}
        \caption{Probability distributions and ratios for $X_\mathrm{SIM}$, $X^*_\mathrm{SIM}$, and $X_\mathrm{DAT}$ for the LHC Olympics model. The $X^*_\mathrm{SIM}$ samples represent the data that results from the trained flow acting upon $X_\mathrm{SIM}$. The good agreement between $X^*_\mathrm{SIM}$ and $X_\mathrm{DAT}$ indicates that the flow has successfully learned to map between the two datasets.}
        \label{fig:LHC_results_SB}
\end{figure*}

In \Figs{fig:LHC_results_sb1}{fig:LHC_results_sb2}, we plot the distributions for $X_\textrm{SIM}$, $X^*_\textrm{SIM}$, and $X_\textrm{DAT}$ for the LHC-like data in SB1 and SB2. For each band, we also plot the ratio of untransformed and transformed simulation distributions to the target distribution. For all features, the transformed simulation $X^*_\textrm{SIM}$ is visually much closer to the target $X_\textrm{DAT}$ than the untransformed simulation.

In \Fig{fig:LHC_results_SR}, we provide the same distributions for $X_\textrm{SIM}$, $X^*_\textrm{SIM}$, and $X_\textrm{DAT}$ in SR. For these plots, we once again see good qualitative agreement between $X^*_\textrm{SIM}$ and $X_\textrm{DAT}$, despite the fact that the flow was not explicitly trained to morph between SR datasets.

In \Tab{tab:LHC_rocs}, we quantify the performance of the flow through the ROC AUC of a binary classifier (with the same architecture as in \Sec{sec:toy_model_results}) trained to discriminate $X^*_\textrm{SIM}$ from $X_\textrm{DAT}$ in each band. In all bands, the AUC is consistent with below 0.51, so our benchmark for indistinguishability of transformed simulation from the target is achieved. 

\begin{table}[]
    \centering
    \begin{tabular}{|c|c|c|}
     \hline
     \hline
  Band & AUC for $X^*_\textrm{SIM}$ vs $X_\textrm{DAT}$ & AUC for $X_\textrm{SIM}$ vs $X_\textrm{DAT}$ \\
        \hline
        \hline
        SB1 & 0.5128 $\pm$  0.0032& 0.5970 $\pm$ 0.0022\\
        \hline
        SR &  0.5034 $\pm$ 0.0010 & 0.5706 $\pm$ 0.0006 \\
        \hline
        SB2 & 0.5153 $\pm$ 0.0054 & 0.5938 $\pm$ 0.0031\\
        \hline
        \hline

    \end{tabular}
    \caption{ROC AUCs for a binary classifier trained to discriminate the transformed simulation $X^*_\textrm{SIM}$ from the target dataset $X_\textrm{DAT}$. For comparison, we also provide the AUCs for a binary classifier trained to discriminate the untransformed simulation $X_\textrm{SIM}$ from the target. }
    \label{tab:LHC_rocs}
\end{table}

\section{Using \textsc{Feta} for anomaly detection}
\label{sec:anomaly_detection}

For a well-trained flow, $X^*_\textrm{SIM}$ should faithfully model the SM background as would be detected at the LHC. However, if the LHC-detected data did contain some anomalous events, this would cause $X^*_\textrm{SIM}$ to differ from $X_\textrm{DAT}$. In this case, an avenue for resonant anomaly detection emerges. We can train a binary classifier to discriminate $X^*_\textrm{SIM}$ in the SR from data in the SR, and a significant deviation found between the distributions might provide evidence of anomalous events in the detected data. This method for anomaly detection relies on the fact that a binary classifier trained to discriminate between two mixed samples (such as a background-only SM template and a detected mixture of SM background and resonant signal) is in fact the optimal classifier for distinguishing pure signal from pure background~\cite{Metodiev:2017vrx,Collins:2019jip,Collins:2018epr}.

\subsection{Signal injection procedure}

To explore the capability of \textsc{Feta} for anomaly detection, we repeat the flow training method as outlined in \Sec{sec:flow_training}. We inject a known number of signal events into the $X_\textrm{DAT}$ (\textsc{Pythia}) dataset. Since our chosen SR extends across the $m_{JJ}$ range [3300, 3700] GeV, this allows for possible detection of the $Z'$ resonance centered at 3500 GeV. We test a range of signal injections, scanning over $n_S \in [300, 500, 750, 1000, 1200, 1500, 2000, 2500, 3000]$ (corresponding to $S/B \approx [0.30\%,\allowbreak 0.49\%,\allowbreak 0.74\%,\allowbreak 0.99\%, \allowbreak1.18\%, 1.48\%,\allowbreak 1.97\%, 2.47\%, 2.96\%]$). As a word of caution: despite the relatively wide SR window chosen, about 20\% of the injected events go into the SB regions.

For each signal injection, we rerun the full \textsc{Feta} pipeline and train a flow to learn the mapping between simulation and data with the injected signal. (This retraining is necessary due to the signal contamination in the sidebands.) We then train a binary classifier (with the same architecture as in \Sec{sec:toy_model_results}) to discriminate transformed simulation from data. 

We compare the results of the \textsc{Feta} method with those from the \textsc{Cathode}, \textsc{Curtains}, and \textsc{Salad} methods. For each alternative method, we use the architecture cited in the respective paper\footnote{Further optimization certainly may yield even better result for the other methods considered, but this optimization task is non-trivial, especially to retain model-agnosticity.}. To place all methods on an equal footing, we use the same set of training and validation dijet events for all methods. Such event breakdowns are given in \Tab{tab:event_breakdown}. Note that all methods use the validation loss to select the best-performing model to be used for further analysis. The \textsc{Feta}, \textsc{Cathode}, and \textsc{Curtains} methods all make use of \textit{oversampling} of the SM background template to achieve better performance, which is a mechanism that allows for the reduction of statistical uncertainties. We investigate the effects of oversampling on the \textsc{Feta} method performance in \App{sec:app_oversampling}.

\begin{table*}[]
    \centering
    \begin{tabular}{|c|c|c|c|c|}
        \hline
        \hline
          Method & Train & Validation (Model Selection) & 
          Template Samples (Evaluation)  &  Oversampling Factor  \\
          \hline
          \hline
          \textsc{Feta} & \parbox[][21pt][c]{100pt}{224k SB simulation \\  186k SB data} & \parbox{100pt}{56k SB simulation \\47k SB data} & 720k SR  & $\times$6  \\ 
           \hline
           \textsc{Cathode} &  186k SB data & 47k SB data & 400k SR & $\times$3 \\
           \hline
         \textsc{Curtains} &  186k SB data & 47k SB data & 1120k SR & $\times$4 \\
         \hline
          \textsc{Salad} & \parbox[][21pt][c]{100pt}{224k SB simulation \\  186k SB data }  & \parbox{100pt}{56k SB simulation \\ 47k SB data} & 120k SR & N/A  \\ 
    \hline
    \hline
    \end{tabular}
    \caption{Number of dijet events used in training and validation, as well as number of events taken as the SR background template for each SM background template construction method. For the \textsc{Curtains} method, SR template samples are generated from transporting both the training and the validation SB samples into the SR. Note that oversampling (the reduction of statistical uncertainties by drawing more SM samples) is not possible with the \textsc{Salad} method, which does not use a generative model for the reference dataset. The oversampling factor is the number of SM background template samples generated over the number of dijet events used to construct the SM background template samples.}
    \label{tab:event_breakdown}
\end{table*}

\begin{figure*}
\begin{subfigure}{.45\linewidth}
\centering
\includegraphics[width=\linewidth]{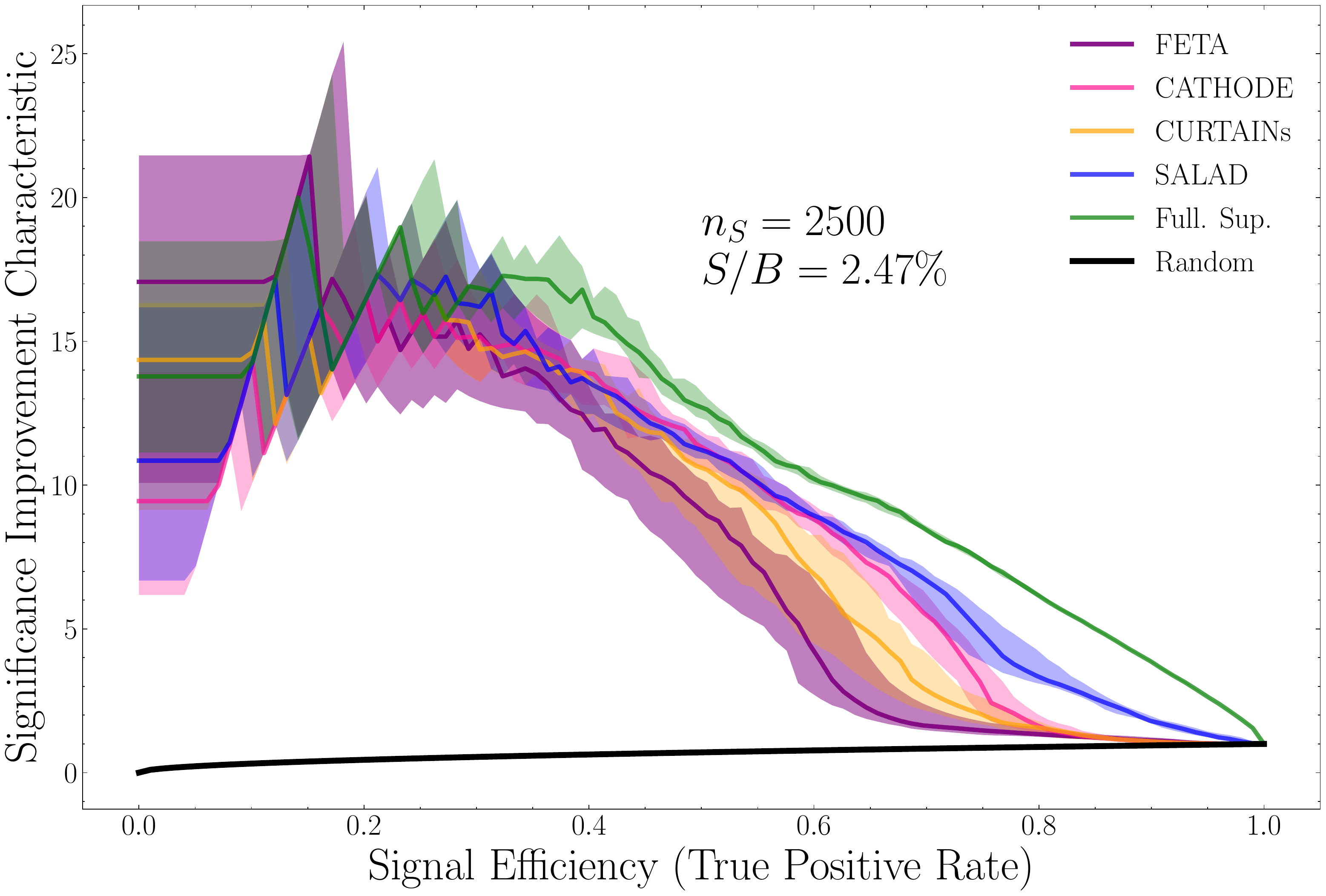}
\caption{Significant improvement characteristic (SIC)}
\label{fig:sic_2500}
\end{subfigure}%
\begin{subfigure}{.45\linewidth}
\centering
\includegraphics[width=\linewidth]{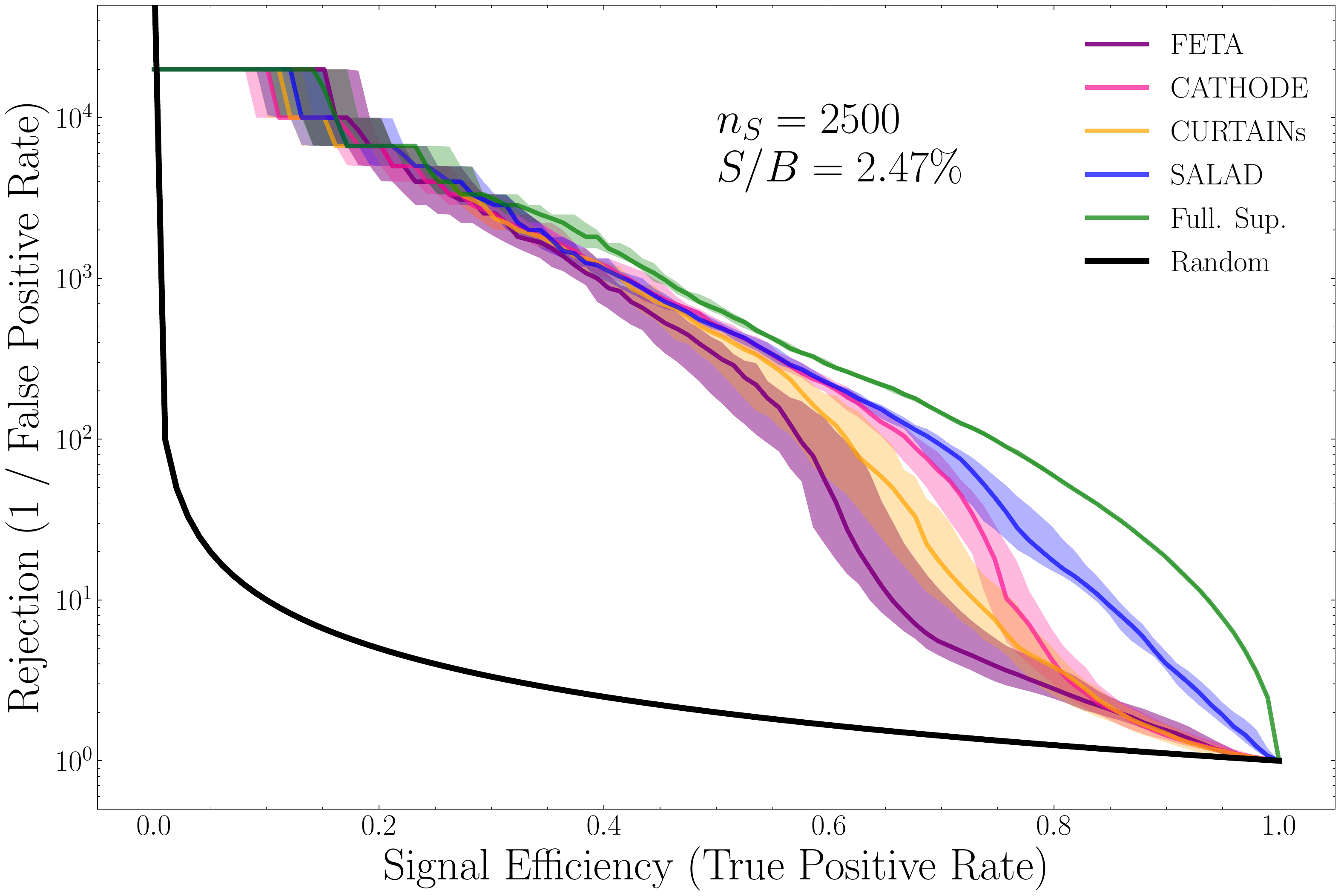}
\caption{Rejection}
\label{fig:rej_2500}
\end{subfigure}\\[1ex]
\begin{subfigure}{\linewidth}
\centering
\includegraphics[width=.45\linewidth]{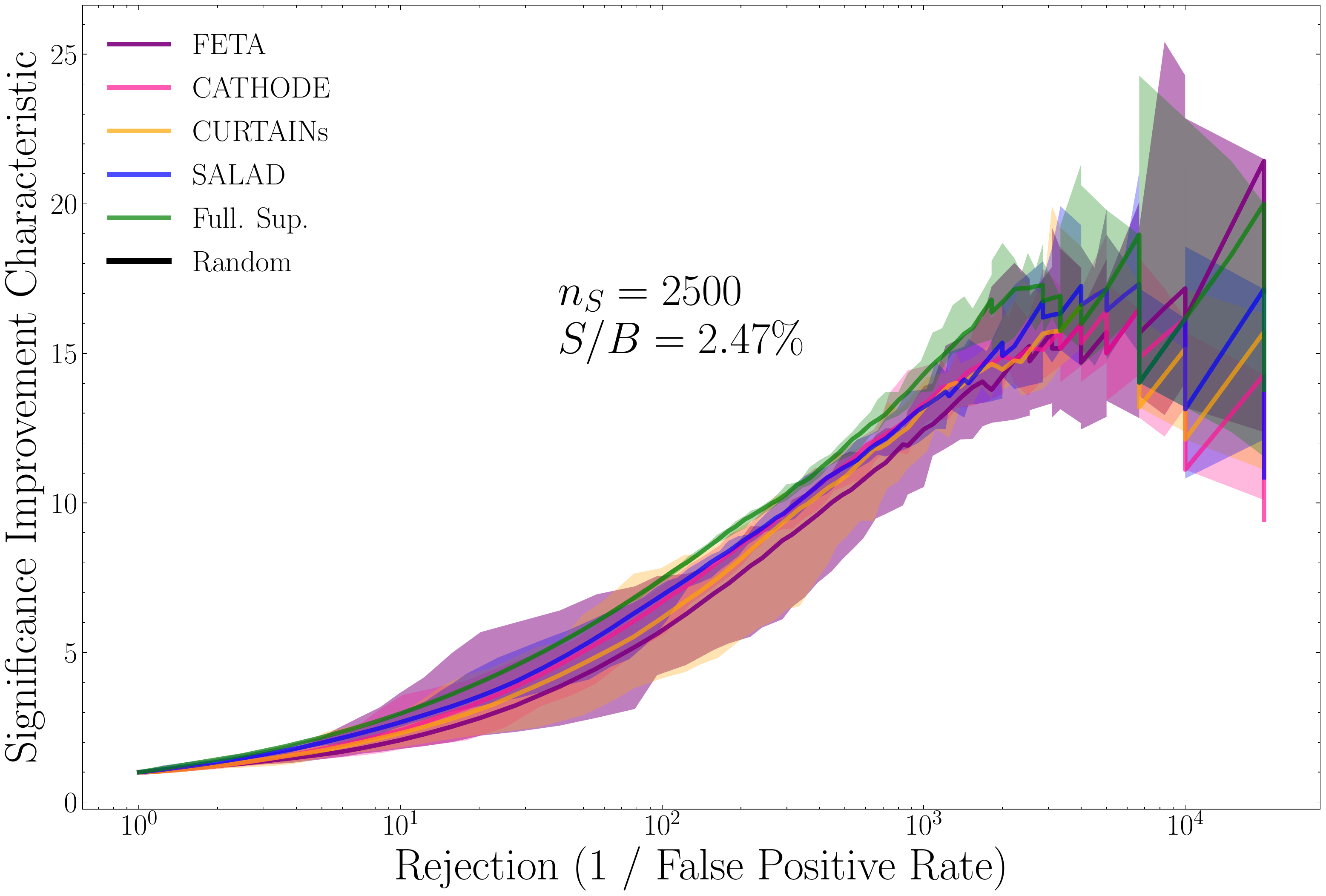}
\caption{SIC against rejection }
\label{fig:sic_vs_rej_2500}
\end{subfigure}
 \caption{Various performance metrics for a binary classifier trained to discriminate a constructed SM background template from detected SR data. We retrain the binary classifier 20 times, each with a different random seed. Curves illustrate the median of these classifier runs, and bands represent the spread across the (16, 84) percentiles. ``Full. Sup." corresponds to a fully supervised classifier trained on pure signal and pure background. }
 \label{fig:ad_results}
\end{figure*}

All methods are evaluated by training a binary classifier to discriminate the SM background template samples from a set of 100k dijet data events from the SR. All classifiers are tested on the same set of 20k signal and 20k background SR dijet events, which were not used at any point during the training or validation procedures.

\subsection{Anomaly detection performance summary}

In \Fig{fig:ad_results}, we show a selection of summary plots from this final binary classifier corresponding to 2500 injected signal events ($S/B = 1.97\%$, $S/\sqrt{B} \approx 7.9$). Each curve represents the mean performance of 20 different random classifiers trained to discriminate the given SM background template from the ``detected" events in the SR. Errorbands represent the spread across the (16, 84) percentiles across these 20 runs. Note that the errorbands are comparable for \textsc{Feta}, \textsc{Cathode}, \textsc{Curtains}, and \textsc{Salad}. 

In \Fig{fig:sic_2500}, we provide the significance improvement characteristic curves, given by SIC = $\frac{\textrm{true positive rate}}{\sqrt{\textrm{false positive rate}}}$. The SIC can be interpreted in the limit of large $S$ and $B$ as the gain in signal significance (i.e. the multiplicative factor) over the initial significance that can be achieved by making a well-motivated cut on the dataset. Therefore for an optimally performing classifier, we expect to see the SIC $>> 1$. In \Fig{fig:rej_2500}, we provide the classifier rejection curves, given by the reciprocal of the false positive rate. An optimally performing classifier should see this rejection also be $>> 1$. Finally, in \Fig{fig:sic_vs_rej_2500}, we plot the SIC curves against the rejection curves. 

In addition to the curves for the four methods considered, we also provide the curves corresponding to a fully supervised classifier, i.e. a classifier trained on perfectly labeled signal and background events. This curve demonstrates the maximum possible performance to discriminate SM physics from anomalous physics. Importantly, the fully supervised classifier should \textit{not} be interpreted as the limiting case of an idealized anomaly detection study, which would come from a classifier trained to discriminate detected data from perfectly simulated background.

We find that the \textsc{Feta} method is competitive with the performance of a fully supervised classifier at signal efficiencies of around 0.3 and lower. This performance is encouraging as in practice, we would expect \textsc{Feta} to be used primarily in this range (for higher signal-efficiencies, the amount of background is sufficiently large that non-fully-supervised methods are unlikely to effectively find the signal). Indeed, the performance curves of \textsc{Feta} for all three metrics, (SIC, rejection, and SIC vs. rejection) very closely align with those of \textsc{Cathode}, which was demonstrated to be state-of-the-art, and \textsc{Curtains}.

\begin{figure}
    \centering
    \includegraphics[width=\linewidth]{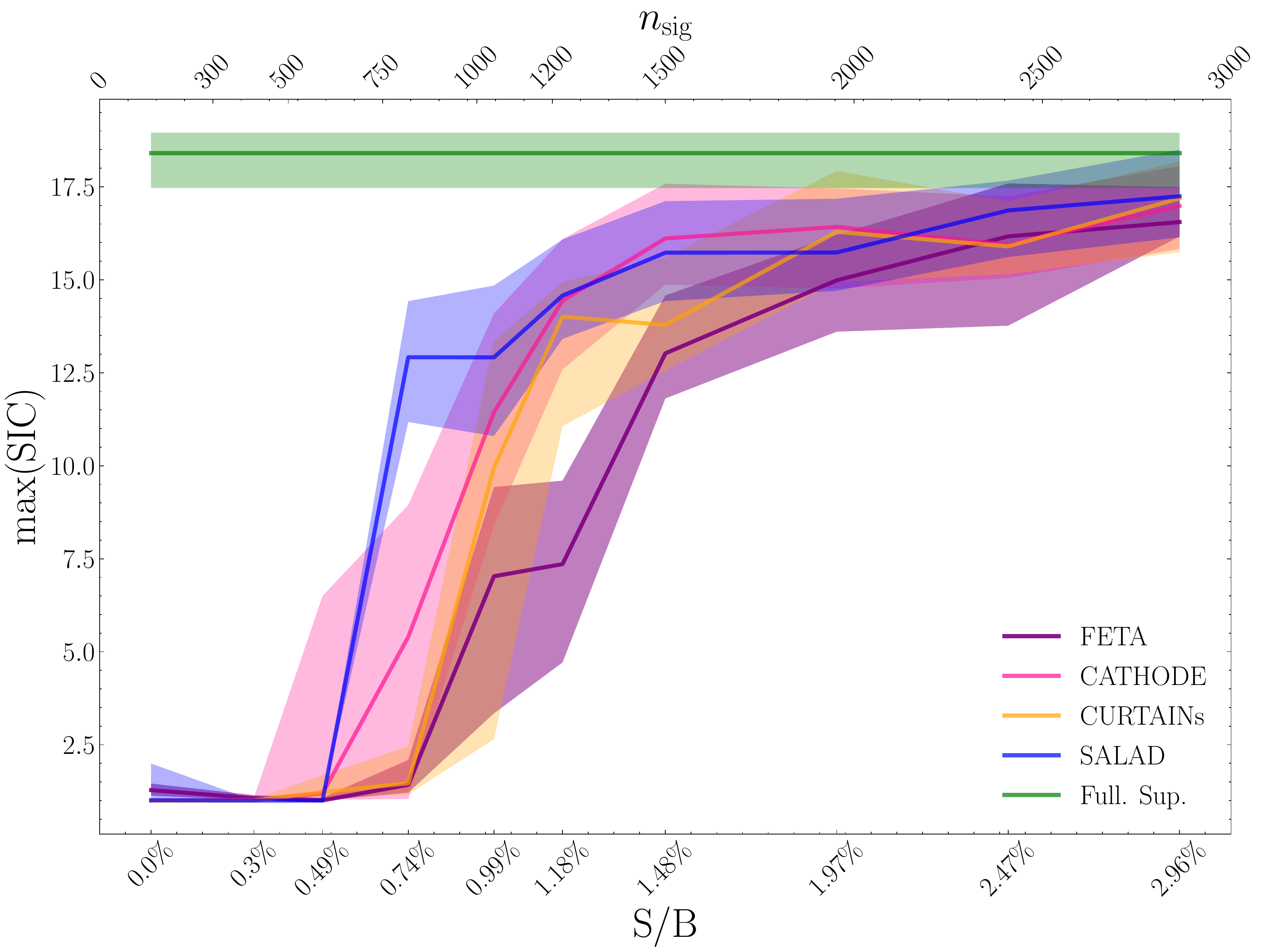}
    \caption{Max significance improvement characteristic (SIC) as a function of the fraction of injected signal events. }
    \label{fig:maxsic}
\end{figure}

In \Fig{fig:maxsic}, for each signal injection, we plot the maximum of the SIC, where the maximum is taken across all signal efficiencies. Since the SIC relates to an increase in signal significance from a well-motivated cut, the max(SIC) is then the best possible cut for a given signal. Across the board, \textsc{Feta}, \textsc{Cathode}, \textsc{Curtains}, and \textsc{Salad} achieve similar performances, all of them becoming just about consistent with a fully supervised classifier at $\sim$1.97\% signal injection. The $S/B$ corresponding to a minimum detectable signal (set to be $\sim$3 for ``observation" of BSM physics) lies between 0.49\% and 0.74\%, as calculated by max(SIC)$\times (S/\sqrt{B})$.

\section{Conclusions and Outlook}
\label{sec:conclusions}

In this study, we have proposed a new method, \textsc{Feta}, for SM background construction that can be used for resonant anomaly detection. The method is simulation-assisted and relies on feature morphing (rather than reweighting) between a reference set of simulated SM data and detected data.

\textsc{Feta} can be seen as a hybrid of \textsc{Curtains} and \textsc{Salad}: like \textsc{Curtains}, \textsc{Feta} uses a flow-based architecture that allows for feature morphing from a reference dataset to construct a SM template. This morphing property performs well in low-density regions of feature space where reweighting methods fail. Like \textsc{Salad}, \textsc{Feta} uses simulated data as the reference dataset. This provides an advantage over data-driven methods as the reference dataset can act as a prior that is free of signal contamination. 

For such simulation-assisted methods of background construction, optimality of transport is desirable as the simulated data provides a physics-informed prior that is expected to be close to the detected data. Therefore an efficient reweighting or morphing function should ideally do as little as possible and reduce to the identity when the simulation is exactly correct. One might ask if the \textit{Transport} flow used in \textsc{Feta} executes the optimal transport between the reference and target datasets, especially as an out-of-the-box normalizing flow contains no loss terms that penalize non-optimal transport. In fact, for the scope of this problem (i.e. morphing between sets of \textsc{Herwig++} and \textsc{Pythia} simulated LHC data), we found that modifications to the flow training method that enforce optimal transport did not significantly change the performance of \textsc{Feta} \cite{Mastandrea:2022vas}. However, these modifications might become important if \textsc{Feta} were applied to morph between a less similar reference and target.

Future avenues for exploration include monitoring the effects of the SB and SR widths on the performance of \textsc{Feta} adding rigorous uncertainty estimates for the performances of all four background construction methods considered in this study, and testing the four methods on a resonant anomaly other than the LHC Olympics one.

\section*{Code availability}

\noindent The code can be found at \href{https://github.com/rmastand/FETA}{https://github.com/rmastand/FETA}.

\begin{acknowledgments}
The authors thank Barry Dillon for his useful comments on the manuscript. The authors also thank Johnny Raine for crafting the acronym for this method.

BN and RM are supported by the U.S. Department of Energy (DOE), Office of Science under contract DE-AC02-05CH11231. TG and SK would like to acknowledge funding through both the SNSF Sinergia grant called ``Robust Deep Density Models for High-Energy Particle Physics and Solar Flare Analysis" (RODEM) with funding number CRSII$5\_193716$, and the SNSF project grant called ``At the two upgrade frontiers: machine learning and the ITk Pixel detector" with funding number 200020\_212127. This material is based upon work supported by the National Science Foundation Graduate Research Fellowship Program under Grant No. DGE 2146752. Any opinions, findings, and conclusions or recommendations expressed in this material are those of the authors and do not necessarily reflect the views of the National Science Foundation.

\end{acknowledgments}

\clearpage

\appendix

\section{Oversampling}
\label{sec:app_oversampling}

In this section, we investigate the effects of oversampling on the performance of the \textsc{Feta} background template. Oversampling was found to greatly improve the performance of the \textsc{Cathode} and \textsc{Curtains} background templates, and similar improvements were found for \textsc{Feta}.

Oversampling refers to the reduction of statistical uncertainties by using a larger number of events. This method is only possible when the background template construction makes use of a density estimator for the reference dataset that can be sampled from multiple times. With \textsc{Feta}, the \textit{Base density} flow (defined in \Sec{sec:data}) plays this role: without oversampling, \textsc{Feta} transports from $X_\textrm{SIM}$ to $X_\textrm{DAT}$; with oversampling, \textsc{Feta} draws samples $X_\textrm{BD}$  from the \textit{Base density} and transforms these to $X_\textrm{DAT}$.

\begin{figure*}[h]
    \begin{subfigure}[b]{\columnwidth}
    \includegraphics[width=\textwidth]{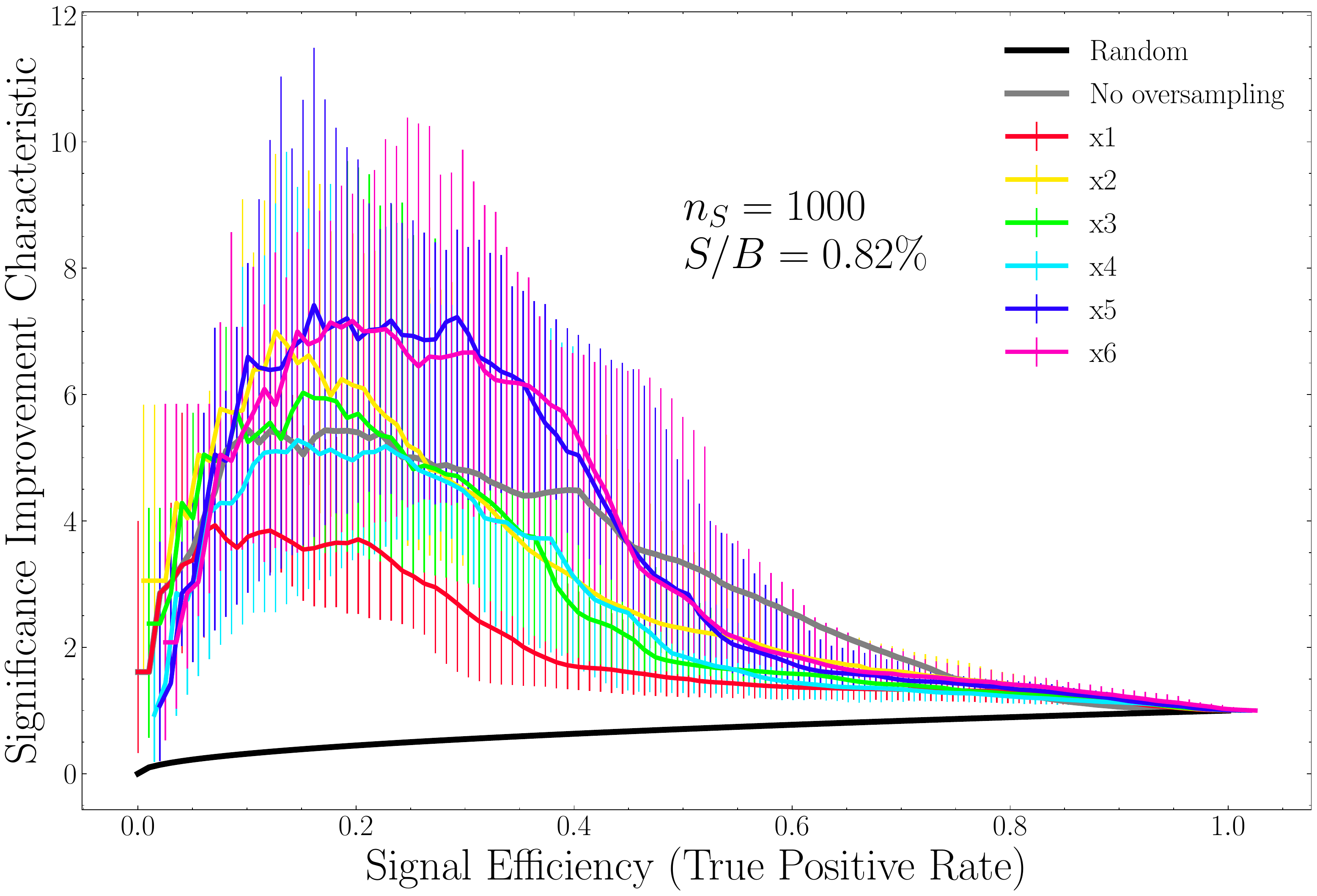}
        \caption{Significance Improvement Characteristic}
    \label{fig:oversampling_sic_1000}
    \end{subfigure}\hfill
    \begin{subfigure}[b]{\columnwidth}
        \includegraphics[width=\textwidth]{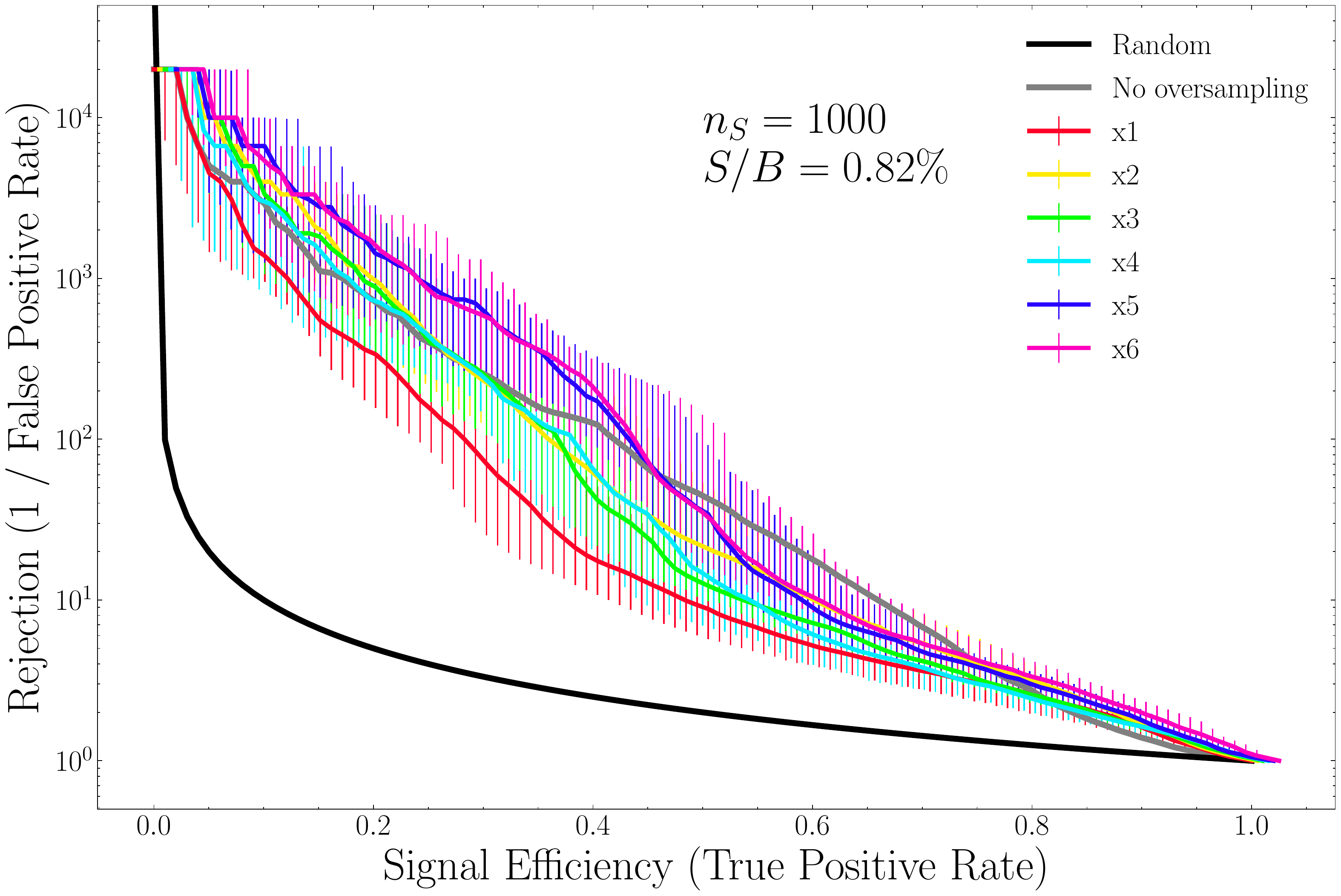}
        \caption{Rejection}
    \label{fig:oversampling_rej_1000}
    \end{subfigure}

    \caption{Performance metrics for a binary classifier trained to discriminate a constructed SM background
template from detected SR data (with 1000 injected signal events) at various factors of oversampling.  We retrain the binary classifier 20 times, each with a different random seed. Curves illustrate the median of these classifier runs, and bands represent the spread across the (16, 84) percentiles.}
    \label{fig:osamp3000}
\end{figure*}

In \Fig{fig:oversampling_sic_1000} and \Fig{fig:oversampling_rej_1000}, we plot the SIC and rejection (respectively) curves corresponding to oversampling factors from one (120k SR events) through six (720k SR events) for a signal injection of 1000 ($S/B = 0.99\%)$. We find that the classifier performance generally increases to a peak at the oversampling factor of $\times$5, after which performance saturates. However, for signal injections of $\geq2000$ ($S/B = 1.97\%$), all oversampling factors perform similarly. 

\section{Signal vs. Background Correlation}
\label{sec:app_correlation}

In this section, we compare in more detail the events produced by the various SM background template construction methods (\textsc{Feta}, \textsc{Cathode}, \textsc{Curtains}, and \textsc{Salad}). 

We expect all of the methods to produce background samples that are significantly different from anomalous events. This is what allows us to detect resonant anomalies when training a classifier to discriminate between SR background samples and SR detected data. However, it is not obvious (or expected) that the background samples will be similar between methods. In fact, we might expect classifier scores to be uncorrelated for background events.

\begin{figure*}
\begin{subfigure}{.8\linewidth}
\centering
\includegraphics[width=\linewidth]{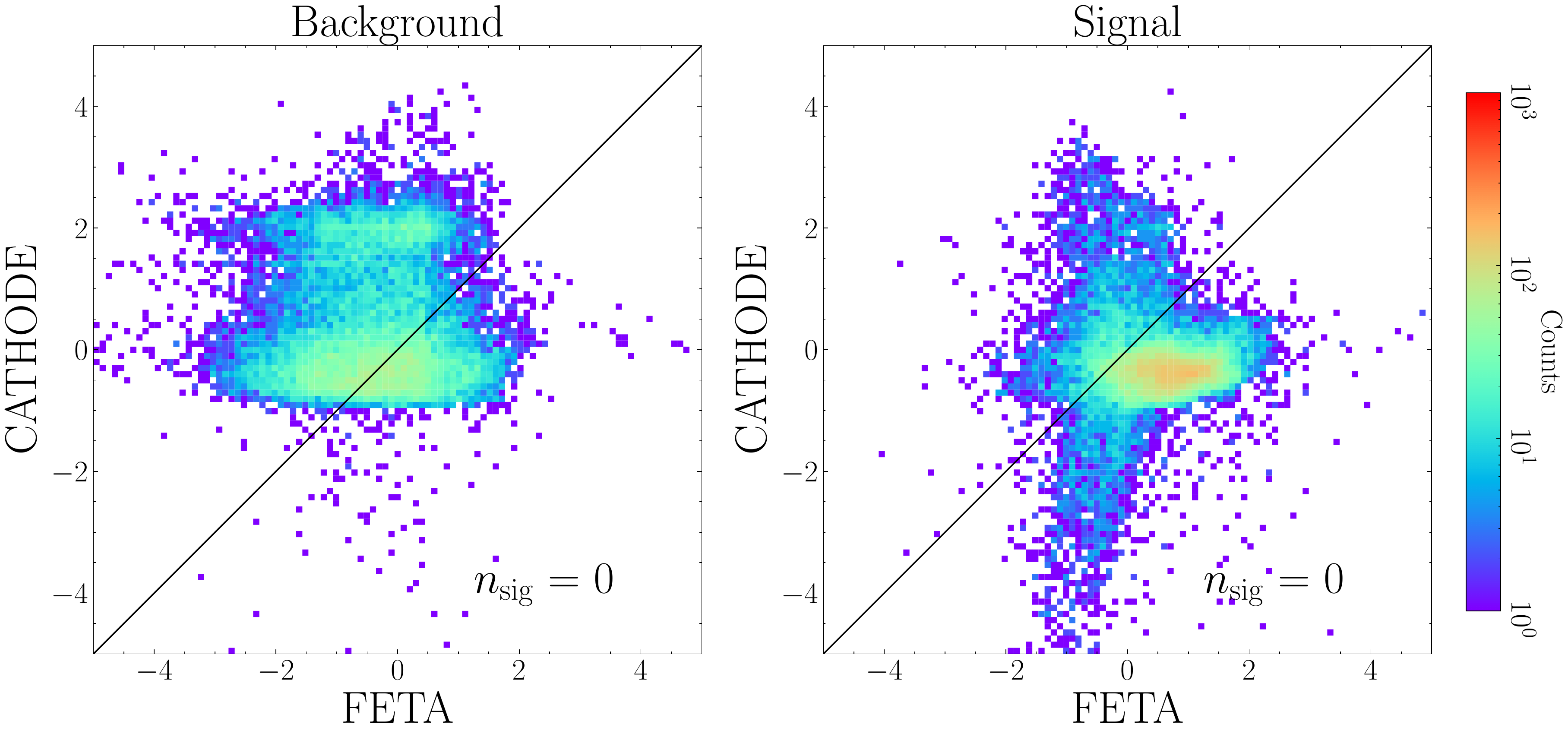}
\caption{\textsc{Feta} vs. \textsc{Cathode}.}
\label{fig:feta_v_cathode0}
\end{subfigure}
\begin{subfigure}{.8\linewidth}
\centering
\includegraphics[width=\linewidth]{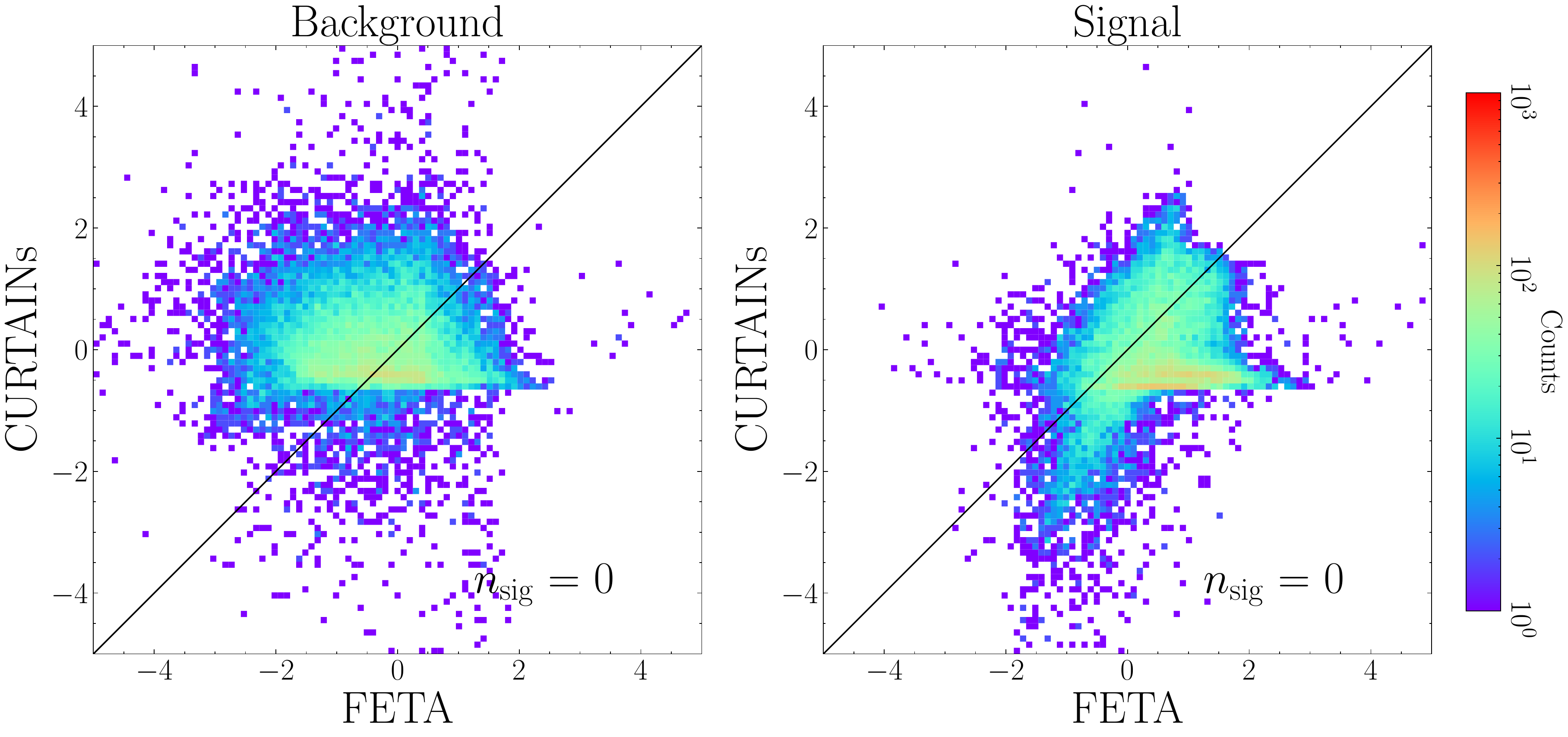}
\caption{\textsc{Feta} vs. \textsc{Curtains}.}
\label{fig:feta_v_curtains0}
\end{subfigure}
\begin{subfigure}{.8\linewidth}
\centering
\includegraphics[width=\linewidth]{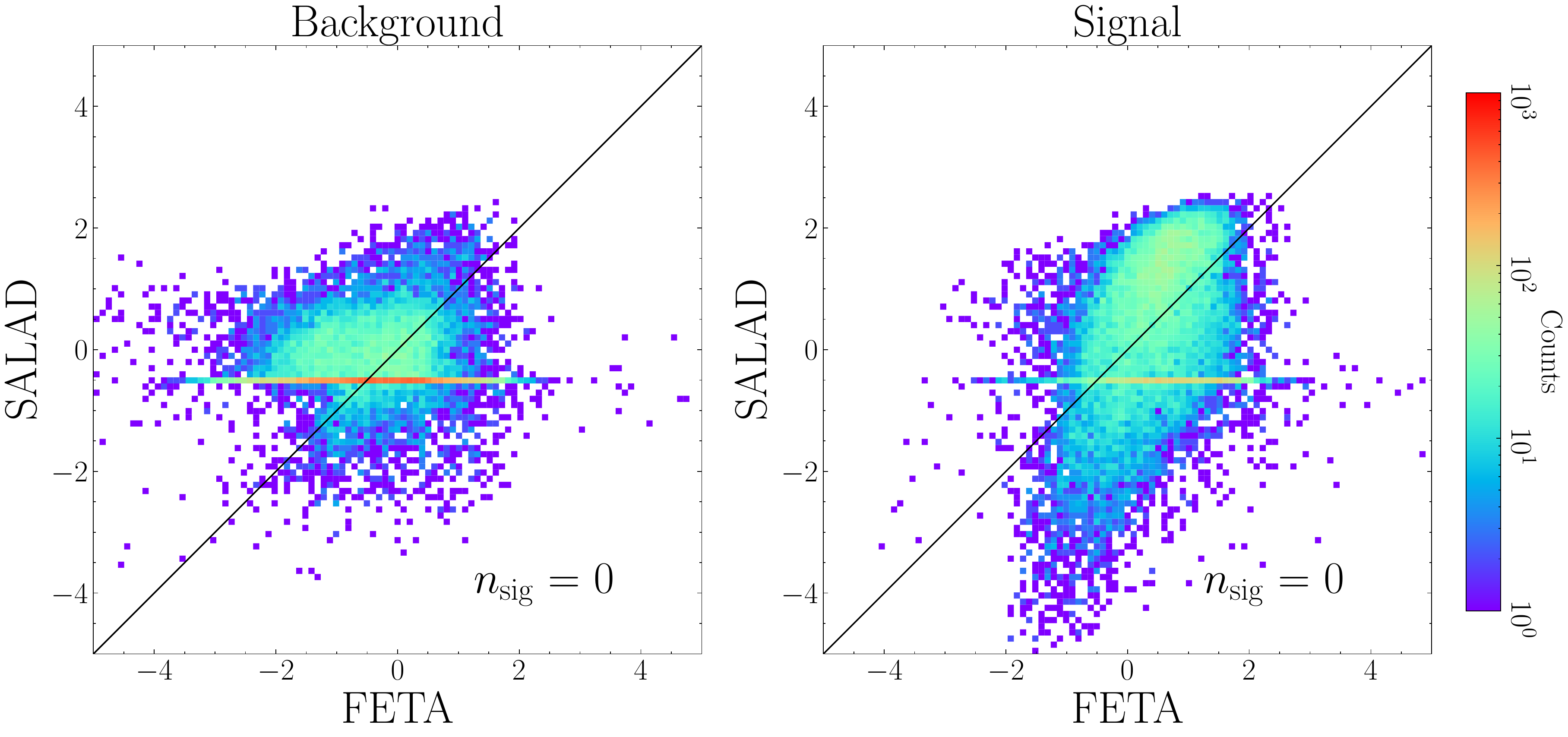}
\caption{\textsc{Feta} vs. \textsc{Salad}.}
\label{fig:feta_v_salad0}
\end{subfigure}
    \caption{Classifier scores for a binary classifier trained to discriminate a given constructed SM background template from detected SR data with no injected signal events and evaluated on pure signal and pure background. The scores have been normalized to have zero mean and unit variance.}
    \label{fig:feta_v_0}
\end{figure*}

\begin{figure*}
  \begin{subfigure}{.8\linewidth}
\centering
\includegraphics[width=\linewidth]{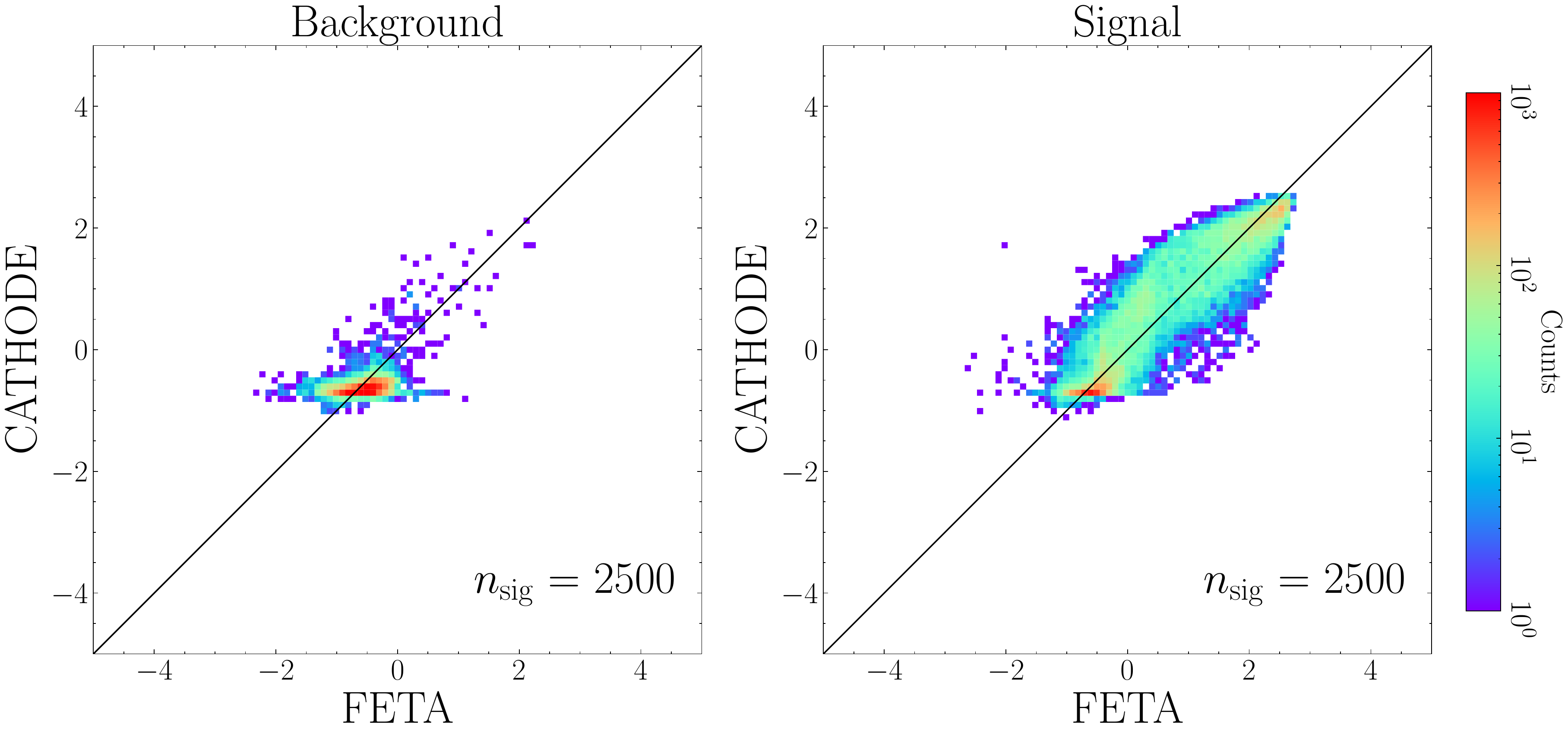}
\caption{\textsc{Feta} vs. \textsc{Cathode}.}
\label{fig:feta_v_cathode2500}
\end{subfigure}
\begin{subfigure}{.8\linewidth}
\centering
\includegraphics[width=\linewidth]{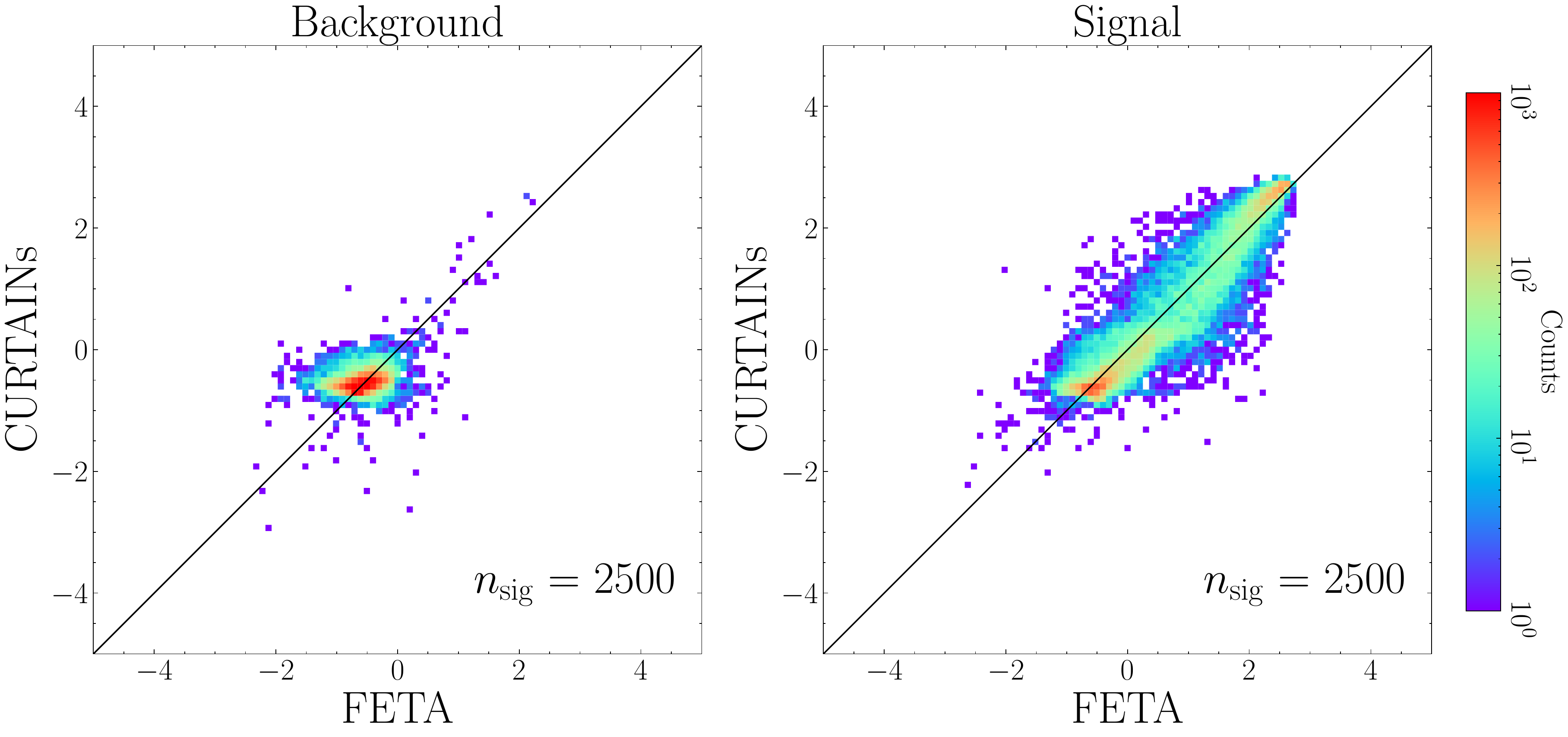}
\caption{\textsc{Feta} vs. \textsc{Curtains}.}
\label{fig:feta_v_curtains2500}
\end{subfigure}
\begin{subfigure}{.8\linewidth}
\centering
\includegraphics[width=\linewidth]{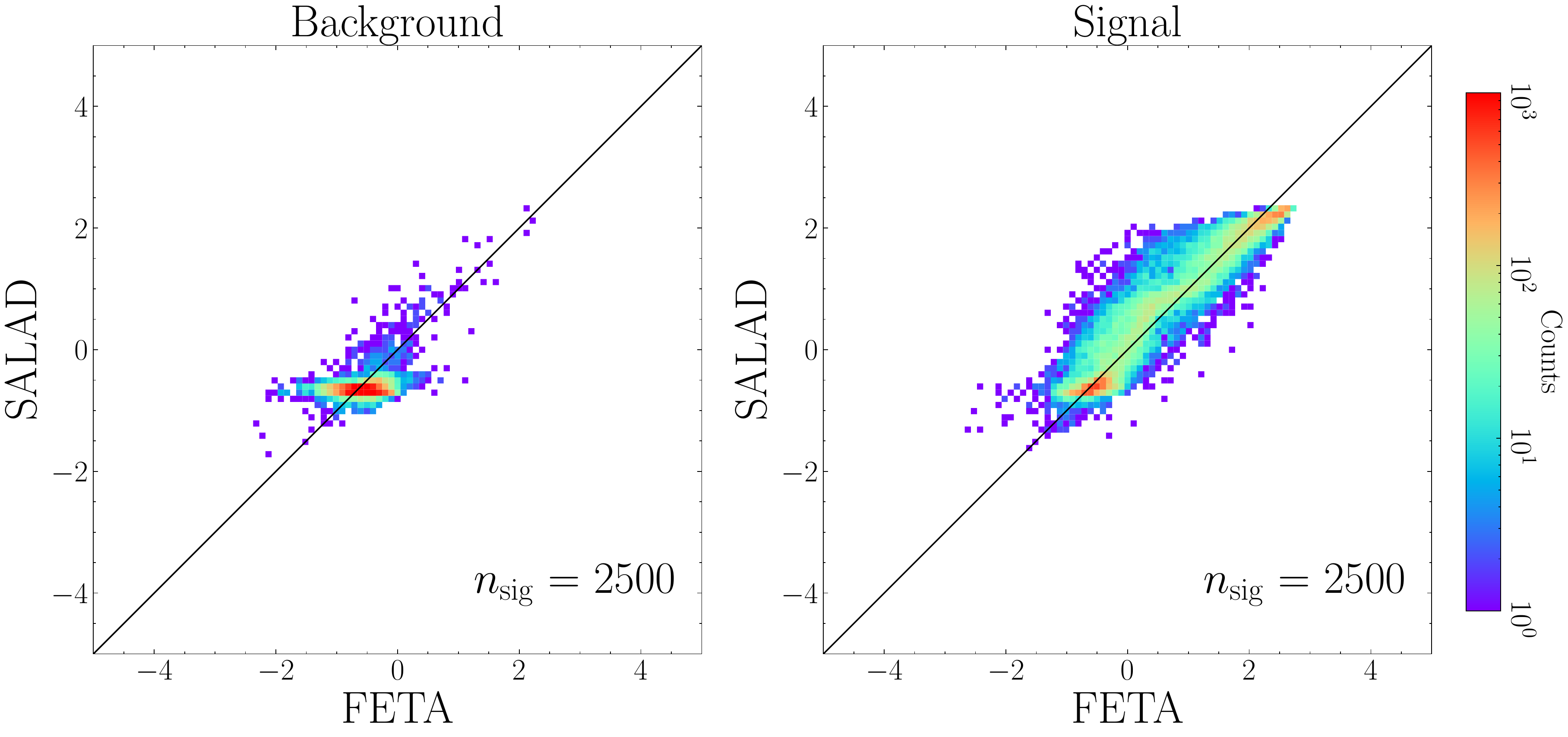}
\caption{\textsc{Feta} vs. \textsc{Salad}.}
\label{fig:feta_v_salad2500}
\end{subfigure}

    \caption{Classifier scores for a binary classifier trained to discriminate a given constructed SM background template from detected SR data with 2500 injected signal events and evaluated on pure signal and pure background. The scores have been normalized to have zero mean and unit variance.}
    \label{fig:feta_v_3000}
\end{figure*}

To probe how different the phase spaces are between construction methods: we train a binary classifier (with the architecture as described in \Sec{sec:toy_model_results}) to discriminate between SR background samples from SR detected data with 0 injected signal events. We evaluate the trained classifier on the test set of 20k signal and 20k background dijet events to get a score between 0 and 1 for each event. (Note that these scores are different than the AUC scores considered in the main analysis.) We finally plot these scores (standardized to zero mean and unit variance) for different methods as a function of each other, for \textsc{Feta} vs. \textsc{Cathode} in \Fig{fig:feta_v_cathode0},  for \textsc{Feta} vs. \textsc{Curtains} in \Fig{fig:feta_v_curtains0}, and for \textsc{Feta} vs. \textsc{Salad} in \Fig{fig:feta_v_salad0}.

For both such cases, we find that the scores assigned to background events derived from classifiers trained on SM background samples from \textsc{Feta} are broadly uncorrelated with those from classifiers trained on SM background samples from the other three methods. There is perhaps a mild degree of correlation between scores for signal events, but the correlation would not be expected to be strong as the classifiers saw no signal events during training.

We then repeat the analysis, this time training the classifiers to discriminate between SR background samples from SR detected data with 2500 injected signal events. We plot the same standardized scores for \textsc{Feta} vs. \textsc{Cathode} in \Fig{fig:feta_v_cathode2500},  for \textsc{Feta} vs. \textsc{Curtains} in \Fig{fig:feta_v_curtains2500}, and for \textsc{Feta} vs. \textsc{Salad} in \Fig{fig:feta_v_salad2500}.

In these cases, there is a large degree of correlation between the classifier scores assigned to signal events between \textsc{Feta} and all of the other construction methods. This result indicates that all of the SM background construction methods are (roughly) equally sensitive to this particular LHCO anomaly.

\bibliography{main,HEPML}

\end{document}